\documentclass[bj,doublespacing,noinfoline]{imsart}

\input{preamble/preamble}
\newcommand{\captitleno}[2]{\caption{\footnotesize Box plots of joint (J) and sequential (S) ML estimates and their difference for sample sizes $N = 500, 2500, 25000$, if the data is generated from $C_{1:3}$ in \autoref{#1} and the pair-copula
families of the \SVCM{} are given by the corresponding \PVCA{}.
The dotted line indicates the  #2 and zero, respectively.
The end of the whiskers is 0.953 times the inter-quartile range, corresponding to approximately 95\%  coverage if the data is generated by a normal distribution.
 }}

\newcommand{\scaleit}{0.48}

\usepackage{amsfonts}
\usepackage{amssymb}
\usepackage{versions}
\excludeversion{showproof}
\newcommand{\condcop}{\text{\LARGE\texthtc}} 
\newcommand{\condden}{\text{\texthtc}}
\renewcommand{\condcop}{\mathbb{C}} 
\renewcommand{\condden}{\text{\scriptsize $\mathbb{C}$}} 
\renewcommand{\condcop}{D} 
\renewcommand{\condden}{d} 
\renewcommand{\condcop}{C} 
\renewcommand{\condden}{c} 
\renewcommand{\cs}{\ps}
\renewcommand{\Mydim}{d}
\newcommand{\tree}[1]{{\cal T}_{#1}}
\newcommand{\allvines}[1]{\mathbf{T}_{#1}}
\newcommand{\treeseq}[1]{\mathbf{T}_{1:#1}}
\newcommand{\mpartreeseq}[1]{\pvc{{\cal T}}_{1:#1}}
\newcommand{\svcs}[1]{#1^{\text{\tiny SVC}}}
\newcommand{\pvcl}[1]{#1_{\text{\tiny PVC}}}
\renewcommand{\idxset}{{\cal I}_{1}^d}
\newcommand{\idxsett}{{\cal I}_{2}^d}
\begin{document}

\begin{frontmatter}
\singlespacing
\title{The partial vine copula: A dependence measure and approximation based on the \SA{}
\thanks{A previous version of this paper was circulated on arXiv under the title ``Simplified vine copula models: Approximations based on the simplifying assumption''.}}
\doublespacing
\runtitle{The partial vine copula}

\begin{aug}
\author{\fnms{Fabian} \snm{Spanhel}\thanksref{a,c1}\ead[label=e1]{spanhel@stat.uni-muenchen.de}}
\and
\author{\fnms{Malte S.} \snm{Kurz}\thanksref{a}\ead[label=e2]{malte.kurz@stat.uni-muenchen.de}}
\address[a]{Department of Statistics, Ludwig-Maximilians-Universit{\"a}t M{\"u}nchen, Akademiestr. 1, 80799 Munich, Germany.}
\printead{e1,e2}
\thankstext{c1}{Corresponding author.}
\runauthor{F. Spanhel and M. S. Kurz}
\end{aug}


\begin{abstract}
\Svcm{s} (\SVCM{s}), or pair-copula constructions, have become an important tool in high-dimensional dependence modeling. 
So far, specification and estimation of \SVCM{s} has been conducted under the simplifying assumption, i.e., all bivariate conditional copulas of the  vine are assumed to be bivariate {unconditional} copulas.
We introduce the partial vine copula (\PVCA{}) which provides a new multivariate dependence measure and which plays a major role in the approximation of multivariate distributions by \SVCM{s}.
The \PVCA{} is a particular \SVCM{} where to any edge a $j$-th order partial copula is assigned and constitutes a multivariate analogue of the bivariate partial copula. 
We investigate to what extent the \PVCA{} describes the dependence structure of the underlying copula. 
We  show that the \PVCA{} does not minimize the Kullback-Leibler divergence from the true copula and that the best approximation satisfying the \SA{} is given by a vine pseudo-copula.
However, under regularity conditions, stepwise estimators of pair-copula constructions converge to the \PVCA{} 
irrespective of whether the \SA{} holds or not. 
Moreover, we elucidate why the \PVCA{} is the best feasible \SVCM{} approximation  in practice. 
\end{abstract}

\end{frontmatter}

\section{Introduction}
Copulas constitute an important tool to model dependence 
\citep{Nelsen2006,Joe1997,McNeil2005}.
While it is easy to construct bivariate copulas, the construction of flexible high-dimensional copulas is a sophisticated problem.
The introduction of simplified vine copulas (\citet{Joe1996}), or pair-copula constructions (\citet{Aas2009b}), has been an enormous advance for high-dimensional dependence modeling. 
Simplified vine copulas are hierarchical structures, constructed upon a sequence of bivariate unconditional copulas, which capture the conditional dependence between pairs of random variables if the \dgp{} satisfies the \SA{}.
In this case, all conditional copulas of the data generating vine collapse to unconditional copulas and the true copula can be represented in terms of a simplified vine copula. 
Vine copula methodology and application have been extensively developed under the \SA{} \citep{Dissmann2013,Grothe2013,Joe2010,Kauermann2013b,Nikoloulopoulos2012}, with studies showing the superiority of simplified vine copula models over elliptical copulas and nested Archimedean copulas (\citet{Aas2009,Fischer2009}).

Although some copulas can be expressed as a simplified vine copula, the \SA{} is not true in general. 
\citet{HobkHaff2010} point out that the simplifying assumption is in general not valid and provide examples of multivariate distributions which do not satisfy the simplifying assumption.
\citet{Stober2013b} show that the Clayton copula is the only Archimedean copula for which the simplifying assumption holds, while the Student-t copula is the only simplified vine copula arising from a scale mixture of normal distributions.
In fact, it is very unlikely that the unknown data generating process satisfies the simplifying assumption in a strict mathematical sense.
As a result, researchers have recently started to investigate new dependence concepts that are related to the simplifying assumption and arise if it does not hold.
In particular, studies on the bivariate partial copula, a  generalization of the partial correlation coefficient%
, have (re-)emerged lately \citep{Bergsma2004,Gijbels2015,Gijbels2015b,Spanhel2015c,Portier2015}. 

We introduce the partial vine copula (\PVCA{}) which constitutes a multivariate analogue of the bivariate partial copula and which generalizes the partial correlation matrix. 
The \PVCA{} is a particular simplified vine copula  where to any edge a $j$-th order partial copula is assigned.
It provides a new multivariate dependence measure for a $d$-dimensional random vector in terms of $d(d-1)/2$ bivariate unconditional copulas and can be readily estimated for high-dimensional data \citep{Nagler2015}. 
We investigate several properties of the \PVCA{} and show to what extent the dependence structure of the underlying distribution is captured. 
The \PVCA{} plays a crucial role in terms of approximating a multivariate distribution by a simplified vine copula (SVC). 
We show that many estimators of \SVCM{s} converge to the \PVCA{} if the \SA{} does not hold.
However, we also prove that the \PVCA{} may not minimize the Kullback-Leibler divergence from the true copula and thus may  not be the best approximation in the space of simplified vine copulas. 
This result is rather surprising, because it implies that it may not be optimal to specify the true copulas in the first tree of a \svca{}.
Moreover, joint and stepwise estimators of \SVCM{s} may not converge to the same probability limit any more if the \SA{} does not hold.
Nevertheless, due to the prohibitive computational burden or simply because only a stepwise model selection and estimation is possible, the \PVCA{} is the best feasible \SVCM{} approximation in practice.
Moreover, the \PVCA{} is used by \citep{Nagler2015} to construct a 
new non-parametric estimator of a multivariate distribution that can outperform classical non-parametric approaches
and by \citep{Kurz2017} to test the simplifying assumption in high-dimensional vine copulas.
All in all, these facts highlight the great practical importance of the \PVCA{} for multivariate dependence modeling.

The rest of this paper is organized as follows.
(Simplified) vine copulas, the simplifying assumption, conditional and partial copulas, are discussed in \autoref{sec_vines_intro}. 
The \PVCA{} and $j$-th order partial copulas are introduced in \autoref{hoppvc}.
Properties of the \PVCA{} and some examples are presented in \autoref{hop_pvc_propex}.
In \autoref{sec_simp_approx_3d} we analyze the role of the \PVCA{} for simplified vine copula approximations and explain why the \PVCA{} is the best feasible approximation in practical applications.
A parametric estimator for the \PVCA{} is presented in \autoref{practice} and implications for the stepwise and joint maximum likelihood estimator of simplified vine copulas are illustrated. 
\autoref{sec_conclusion} contains some concluding remarks.

The following notation and assumptions are used throughout the paper. 
We write $\Myx_{1:\Mydim}:= (\Myx_1,\ldots,\Myx_\Mydim)$, so that
$
F_{X_{1:d}}(x_{1:d}) :=
\P(\forall i=1,\ldots,d\colon X_i\leq x_i)$, and
$\text{d}x_{1:d}:= \text{d}x_1\ldots \text{d}x_d$ to denote the variables of integration in 
$\int f_{X_{1:d}}(x_{1:d})\text{d}x_{1:d}$. 
$C^\perp$ refers to the independence copula. $X \perp Y$ means that $X$ and $Y$ are stochastically independent.
For $1\leq k\leq d$, the partial derivative of $g$ w.r.t. the $k$-th argument is denoted by $\partial_kg(x_{1:d})$.
We write $\ivb{A}=1$ if $A$ is true, and $\ivb{A}=0$  otherwise.
For simplicity, we assume that all random variables are real-valued and continuous.
In the following, let $d\geq 3$, if not otherwise specified, and ${\cal C}_d$ be the space of absolutely continuous $d$-dimensional copulas with  positive density (a.s.). 
The distribution function of a random vector $U_{1:d}$ with uniform margins is denoted by $F_{1:d}=C_{1:d}\in {\cal C}_d$.
We set ${\cal I}_{l}^d:= \{(i,j)\colon j=l,\ldots,d-1,i=1,\ldots,d-j\}$
and 
$\condset:=i+1:i+j-1:=i+1,\ldots,i+j-1$.
We focus on D-vine copulas, but all results carry over to regular vine copulas  (\citet{Bedford2002}, \citet{Kurowicka2011}). 
An overview of the used notation can be found in \autoref{table_notation}.
{All proofs are deferred to the appendix.}

\begin{table}[h!]
\footnotesize
\centering
\caption{
Notation for simplified D-vine copulas. $U_{1:d}$  has standard uniform margins, $d\geq 3,(i,j)\in \idxset, k=i,i+j$.
}
\label{table_notation}
\renewcommand{\arraystretch}{1.7}
\begin{tabular}{c|p{13.2cm}}
Notation & Explanation
\\
\hline
$F_{1:d}$ or $C_{1:d}$ &  cdf and copula of $U_{1:d}$
\\
\hline
${\cal C}_d$ & space of $d$-dimensional copulas with  positive density
\\
\hline
$\SPCC{C}_d$ & space of $d$-dimensional simplified D-vine copulas with  positive density
\\ \hline
$\idxset$ &
$\idxset:=\{(i,j)\colon j=1,\ldots,d-1,i=1,\ldots,d-j\}$, 
the conditioned set of a D-vine copula density
\\ \hline
$\condset$ &
$\condset := i+1:i+j-1:= i+1,\ldots,i+j-1$, the conditioning set of an edge in a D-vine
\\
\hline
$U_{k|\condset}$ & $F_{k|\condset}(U_k|U_{\condset})$, conditional 
probability 
integral transform (CPIT) of $U_k$ w.r.t. $U_{\condset}$
\\
\hline 
$\condcop_{i,i+j\cs \condset}$
& bivariate conditional copula of $F_{i,{i+j}|{\condset}}$, i.e.,
\mbox{$\condcop_{i,i+j\cs \condset} = F_{U_{i|\condset},U_{i+j|\condset}|U_{\condset}}$
}
\\
\hline
$\svcs{C}_{i,i+j\ps \condset}$ 
& \mbox{arbitrary bivariate (unconditional) copula that is used to model}
 $\condcop_{i,i+j\cs \condset}$
\\
\hline
$\parsign{C}_{i,i+j\ps \condset}$
& partial copula of  $\condcop_{i,i+j\cs \condset}$, i.e.,
$\parsign{C}_{i,i+j\ps \condset} = F_{U_{i|\condset},U_{i+j|\condset}}$
\\
\hline
$\pvc{C}_{i,i+j\ps \condset}$ &
$(j-1)$-th order partial copula of $\condcop_{i,i+j\cs \condset}$
\\
\hline
$\pvc{U}_{k|\condset}$ &
$\pvc{F}_{k|\condset}(U_k|U_{\condset})$, $(j\!-\!2)$-th order partial probability integral transform (PPIT) of $U_k$ w.r.t. $U_{\condset}$ 
\\
\hline
$\pvc{C}_{1:d}$ & 
\mbox{Partial vine copula (PVC) of  $C_{1:d}$, if $d=3$, then }
\mbox{$\pvc{c}_{1:3}(u_{1:3})= c_{12}(u_1,u_2)\ \!
c_{23}(u_2,u_3)\ \!
\pvc{c}_{13\ps 2}(u_{1|2},u_{3|2})$ 
}
\end{tabular}

\end{table}

\section{Simplified vine copulas, conditional copulas, and higher-order partial copulas}
\label{sec_vines_intro}
In this section, we discuss (simplified) vine copulas and  the simplifying assumption.
Thereafter, we introduce the partial copula which can be considered as a generalization of the partial correlation coefficient and as an approximation of a bivariate conditional copula. 

\begin{mydef}
[Simplified D-vine copula or pair-copula construction  -- \citet{Joe1996,Aas2009b}]
\label{sdvinedef}
For   $(i,j)\in\idxset$, let $\svcs{C}_{i,i+j\ps \condset}\in {\cal C}_2$ with density $\svcs{c}_{i,i+j\ps \condset}$.
For $j=1$ and $i=1,\ldots,d-j$, we set $\svcs{C}_{i,i+j\ps \condset} = \svcs{C}_{i,i+1}$ 
{and  $\svcs{u}_{k|\condset} = u_k$  for $ k=i,i+j$.}
For $(i,j)\in \idxsett$, define
\begin{align*}
\svcs{u}_{i| \condset}  := \svcs{F}_{i|\condset}(u_i|u_{\condset}) &= 
\partial_2\svcs{C}_{i,i+j-1\ps \sarg{i}{j-1}}
(\svcs{u}_{i|\sarg{i}{j-1}},
\svcs{u}_{i+j-1|\sarg{i}{j-1}}
),
\\
\svcs{u}_{i+j|\condset} := \svcs{F}_{i+j|\condset}(u_{i+j}|u_{\condset}) &= 
\partial_1\svcs{C}_{i+1,i+j\ps \sarg{i+1}{j-1}}
(\svcs{u}_{i+1|\sarg{i+1}{j-1}},
\svcs{u}_{i+j|\sarg{i+1}{j-1}}
)
.
\end{align*}
Then
\begin{align*}
\svcs{c}_{1:\Mydim}(u_{1:\Mydim}) & = 
\prod_{(i,j)\in \idxset} \svcs{c}_{i,i+j\ps \condset}
\big( \svcs{u}_{i|\condset},\svcs{u}_{i+j|\condset}\big)
\end{align*}
is the density of a $d$-dimensional simplified D-vine copula $\svcs{C}_{1:d}$. 
We denote the space of $d$-dimensional simplified D-vine copulas by $\SPCC{C}_d$.
\end{mydef}
\begin{figure}
\newcommand{\mys}{0.5}
\centering
\begin{subfigure}[b]{\mys\textwidth}
\centering
  \includegraphics{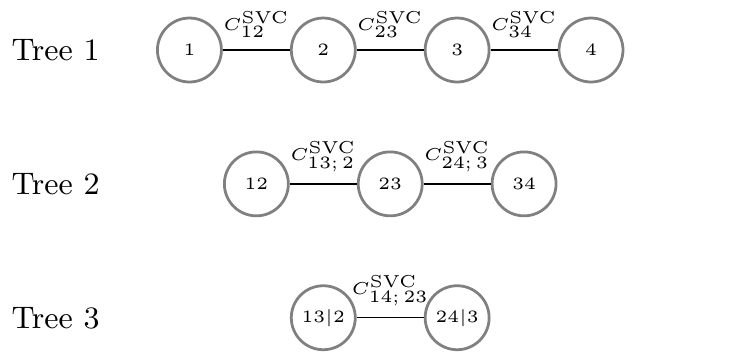}
  \label{svc}
  \caption{Simplified D-vine copula.}
  \end{subfigure}%
\begin{subfigure}[b]{\mys\textwidth}
  \centering
  \includegraphics{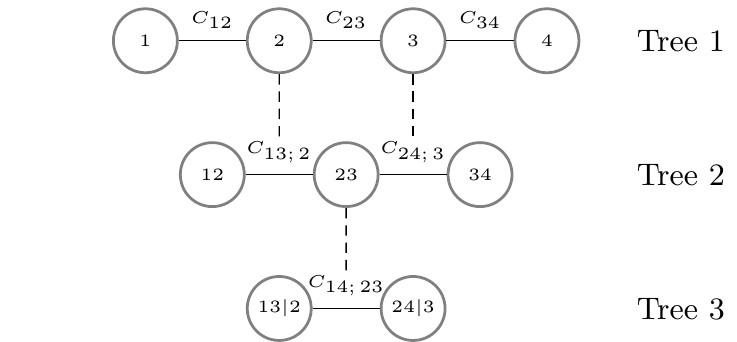}
  \label{vc}
  \caption{D-vine copula.}
\end{subfigure}

\caption{(Simplified) D-vine copula representation if $d=4$. The influence of conditioning variables on the conditional copulas is indicated by dashed lines.}
\label{fig_vine}
\end{figure}

From a graph-theoretic point of view, simplified (regular) vine copulas  can be considered as an ordered sequence of trees, where $j$ refers to the number of the tree and  a bivariate {unconditional} copula $\svcs{C}_{i,i+j\ps\condset}$ is assigned to each of the $d-j$ edges of tree $j$ (\citet{Bedford2002}).
{
The left hand side of \autoref{fig_vine} shows the graphical representation of a simplified D-vine copula for $d=4$, i.e., 
\begin{align*}
\svcs{c}_{1:4}(u_{1:4}) & = 
\underbrace{\svcs{c}_{12}(u_{1},u_2)\svcs{c}_{23}(u_2,u_3)\svcs{c}_{34}(u_3,u_4)}
_{\text{first tree}}
\times
\underbrace{\svcs{c}_{13\ps 2}(\svcs{u}_{1|2},\svcs{u}_{3|2})\svcs{c}_{24\ps 3}(\svcs{u}_{2|3},\svcs{u}_{4|3})}_{\text{second tree}}
\times 
\underbrace{\svcs{c}_{14\ps 2:3}(\svcs{u}_{1|2:3},\svcs{u}_{4|2:3})}_
{\text{third tree}}.
\end{align*}
}%
The bivariate unconditional copulas $\svcs{C}_{i,i+j\ps \condset}$ are also called pair-copulas, so that the resulting model is often termed a pair-copula construction (PCC). 
By means of simplified vine copula models one can construct a wide variety of flexible multivariate copulas because each of the $d(d-1)/2$ bivariate unconditional copulas $\svcs{C}_{i,i+j\ps \condset}$ can be chosen arbitrarily and the resulting model is always a valid $d$-dimensional copula. 
Moreover, a pair-copula construction does not suffer from the curse of dimensions because it is build upon a sequence of bivariate unconditional copulas which renders it very attractive for high-dimensional applications.
Obviously, not every multivariate copula can be represented by a simplified vine copula.
However, every copula can be represented by the following (non-simplified) D-vine copula.

\begin{mydef}
[D-vine copula   --  \citet{Kurowicka2006}]
\label{dvinedef}
Let $U_{1:d}$ be a random vector with cdf $F_{1:d} = C_{1:d}\in {\cal C}_d$.
For $j=1$ and $i=1,\ldots,d-j$, we set ${C}_{i,i+j\ps \condset} = {C}_{i,i+1}$ 
{and  ${u}_{k|\condset} = u_k
$  for $ k=i,i+j$.}
For $(i,j)\in\idxsett$, let $\condcop_{i,i+j\cs \condset}$ denote the conditional copula of $F_{i,i+j|\condset}$
(\autoref{def_conditional_copula}) and let
 $u_{k|\condset}:= F_{k|\condset}(u_k|u_{\condset})$ for  $k=i,i+j$.
The density of a D-vine copula decomposes the copula density of $U_{1:d}$
into $\Mydim(\Mydim-1)/2$ bivariate conditional copula densities 
$\condden_{i,i+j\cs \condset}$ according to the following factorization:
\begin{align*}
c_{1:\Mydim}(u_{1:\Mydim}) & = 
\prod_{(i,j)\in \idxset} \condden_{i,i+j\cs \condset}
( u_{i|{\condset}},u_{i+j|\condset}|u_{\condset}).
\end{align*}
\end{mydef} 

Contrary to a simplified D-vine copula in \autoref{sdvinedef}, a bivariate conditional copula $\condcop_{i,i+j\cs\condset}$, which is in general a function of $j+1$ variables, is assigned to each edge of a  D-vine copula in \autoref{dvinedef}.
The influence of the conditioning variables on the conditional copulas is illustrated by dashed lines in the right hand side of \autoref{fig_vine}.
In applications, the simplifying assumption is typically imposed, i.e., 
it is assumed that all bivariate conditional copulas of the data generating vine copula degenerate to bivariate unconditional copulas.

\begin{mydef}
[The simplifying assumption -- \citet{HobkHaff2010}]
\label{simplidef}
{The D-vine copula in \autoref{dvinedef} satisfies the \SA{} if 
$\condden_{i,i+j\cs \condset}
( \cdot,\cdot|u_{\condset})$ does not depend on $u_{\condset}$
for all $(i,j)\!\in\! \idxsett$.}
\end{mydef}

If the data generating copula satisfies the simplifying assumption, it can be represented by a simplified vine copula, resulting in fast and simple statistical inference. 
Several methods for the consistent specification and estimation of pair-copula constructions have been  developed under this assumption (\citet{HobkHaff2013}, \citet{Dissmann2013}).
However, in view of \autoref{dvinedef} and \autoref{sdvinedef} it is evident that it is extremely unlikely that the data generating vine copula strictly satisfies the \SA{} in practical applications.

Several questions arise if the \dgp{} does not satisfy the \SA{} and a simplified D-vine copula model (\autoref{sdvinedef}) is used to approximate a general D-vine copula (\autoref{dvinedef}).
First of all, what bivariate unconditional copulas $\svcs{C}_{i,i+j\ps\condset}$ should be chosen in \autoref{sdvinedef} to model the bivariate conditional copulas $\condcop_{i,i+j\cs \condset}$ in \autoref{dvinedef} so that the best approximation w.r.t. a certain criterion is obtained? 
What simplified vine copula model do established stepwise procedures (asymptotically) specify and estimate if the \SA{} does not hold for the data generating vine copula?
What are the properties of an optimal approximation?
Before we address these questions in \autoref{sec_simp_approx_3d}, 
it is useful to recall the definition of the conditional and partial copula in the remainder of this section and to introduce and investigate the partial vine copula in \autoref{hoppvc} and \autoref{hop_pvc_propex} because it plays a major role in the approximation of copulas by simplified vine copulas.


\label{sec_partial}
\begin{mydef}[Conditional probability integral transform (CPIT)]
\label{def_cpit}
Let $U_{1:d}\sim F_{1:d}\in {\cal C}_d$, $(i,j)\in \idxsett$ and 
$k=i,i+j$. We call $U_{k|\condset} := F_{k|\condset}(U_k|U_{\condset})$ the conditional probability integral transform of $U_k$ w.r.t. $U_{\condset}$.
\end{mydef}

{It can be readily verified that, under the assumptions in \autoref{def_cpit},
$
U_{k|\condset}\sim {\cal U}(0,1)$ and \mbox{$
U_{k|\condset}\perp U_{\condset}$}. }
Thus, applying the random transformation 
$F_{k|\condset}(\cdot|U_{\condset})$ to $U_k$ removes possible
dependencies between $U_k$ and $U_{\condset}$ {and $
U_{k|\condset}$ can be interpreted as the remaining variation in $U_k$ that can not be explained by $U_{\condset}$}.
This interpretation of the CPIT is crucial for understanding  the conditional and partial copula which are related to the (conditional) joint distribution of CPITs.
The conditional copula has been introduced by \citet{Patton2006b} and we restate its definition here.%
\footnote%
{
Patton's notation for the conditional copula is given by $C_{i,i+j|\condset}$. Originally, this notation has also been used in the vine copula literature \citep{Aas2009b,Kurowicka2011,Acar2012}. 
However, the current notation for a(n) (un)conditional copula that is assigned to an edge of a vine is given by $C_{i,i+j\ps \condset}$ and $C_{i,i+j|\condset}$ is used to denote $F_{U_i,U_{i+j}| U_{\condset}} $
\citep{Joe2010,Stober2013b,Krupskii2013b}.
In order to avoid possible confusions, we use $\condcop_{i,i+j\cs \condset}$ to denote a conditional copula and $\svcs{C}_{i,i+j\ps \condset}$ to denote an unconditional copula. 
}
\newcommand{\Myya}{Y_1}
\newcommand{\Myza}{Y_2}
\newcommand{\myya}{y_1}
\newcommand{\myza}{y_2}
\newcommand{\Mywa}{Z}
\newcommand{\mywa}{z}
{
\newcommand{\Pitu}{U_1(Z)}
\newcommand{\Pitv}{U_2(Z)}
\renewcommand{\Myya}{i}
\renewcommand{\Myza}{{i+j}}
\renewcommand{\myya}{u_i}
\renewcommand{\myza}{u_{i+j}}
\renewcommand{\Mywa}{{\condset}} 
\renewcommand{\mywa}{u_{\condset}} 
\renewcommand{\Pitu}{U_i}
\renewcommand{\Pitv}{U_{i+j}}
\renewcommand{\Pitu}{U_{i|\condset}}
\renewcommand{\Pitv}{U_{i+j|\condset}}
\newcommand{\evala}{a}
\newcommand{\evalb}{b}
\begin{mydef}
[Bivariate conditional copula -- \citet{Patton2006b}] 
\label{def_conditional_copula}
Let $U_{1:d}\sim F_{1:d}\in{\cal C}_d$ and $(i,j)\in \idxsett$.  
The (\as) unique conditional copula $\condcop_{i,i+j\cs \condset}$ of the conditional distribution $F_{\Myya,\Myza|\Mywa}$  is defined by 
\begin{align*}
\condcop_{i,i+j\cs \condset}(\evala,\evalb|\mywa)  &:=
\P(\Pitu \leq \evala,\Pitv \leq \evalb|U_\Mywa = \mywa)
\\& \phantom{:}
 =  F_{\Myya,\Myza|\condset}(F_{\Myya|\Mywa}^{-1}(a|\mywa),
F_{\Myza|\Mywa}^{-1}(b|\mywa)|\mywa).
\end{align*}
\end{mydef}

Equivalently, we have that
\begin{align*} 
F_{\Myya,\Myza|\Mywa}(\myya,\myza|\mywa) = \condcop_{i,i+j\cs \condset}
(F_{\Myya|\Mywa}(\myya|\mywa),F_{\Myza|\Mywa}(\myza|\mywa)|\mywa),
\end{align*}
 so that the effect of a change in $\mywa$ on the conditional distribution
$F_{\Myya,\Myza|\Mywa}(\myya,\myza|\mywa)$ can be separated into two effects.
First, the values of the CPITs, $(F_{\Myya|\Mywa}(\myya|\mywa),F_{\Myza|\Mywa}(\myza|\mywa))$, at which the conditional copula is evaluated,
may change.
Second, the functional form of the conditional copula 
$\condcop_{i,i+j\cs \condset}(\cdot,\cdot|\mywa)$ may vary.
In comparison to the conditional copula, which is the conditional distribution of two CPITs, the partial copula is the unconditional distribution and copula of  two CPITs. 

\begin{mydef}
[Bivariate partial copula - \citet{Bergsma2004}]
\label{def_partial_copula}
Let $U_{1:d}\sim F_{1:d}\in{\cal C}_d$ and $(i,j)\in \idxsett$.  
The partial copula  $\parsign{C}_{i,i+j\ps \condset}$ of the distribution
$F_{\Myya,\Myza|\Mywa}$ is 
defined by 
\begin{align*}
\parsign{C}_{i,i+j\ps \condset}(a,b) & := \mathbb{P}(\Pitu\leq a, \Pitv\leq b).
\end{align*}
\end{mydef}

Since $\Pitu\perp U_\Mywa$ and $\Pitv\perp U_\Mywa$, the partial copula represents the distribution of  random variables which are individually independent of the conditioning vector $U_\Mywa$. This is similar to the partial correlation coefficient, which is the correlation of two random variables from which  the linear influence of the conditioning vector has been removed.
The partial copula can also be interpreted as the expected conditional copula,

\begin{align*}
\parsign{C}_{i,i+j\ps \condset}(a,b) = \int_{\R^{j-1}} 
 \condcop_{i,i+j\cs \condset}(a,b|\mywa)
\text{d}F_{\Mywa}(\mywa),
\end{align*}
and be considered as an approximation of the conditional copula. 
Indeed, it is easy to show that the partial copula $\parsign{C}_{i,i+j\ps \condset}$ minimizes the Kullback-Leibler divergence from the conditional copula $ \condcop_{i,i+j\cs \condset}$ in the space of absolutely continuous bivariate distribution functions.
The partial copula is first  mentioned by \citet{Bergsma2004} who applies the partial copula to test for conditional independence. 
Recently, there has been a renewed interest in the partial copula. 
\citet{Spanhel2015c} investigate properties of the partial copula and mention some explicit examples whereas \citet{Gijbels2015,Gijbels2015b} and \citet{Portier2015} focus on the non-parametric estimation of the partial copula.
}

\section{Higher-order partial copulas and the \pvca{}}
\label{hoppvc}
\newcommand{\sargg}{S_{1K}}
A generalization of the partial correlation coefficient that is different from the partial copula is given by the higher-order partial copula.
To illustrate this relation, let us recall the common definition of the partial correlation coefficient. 
Assume that all univariate margins of $Y_{1:d}$ have zero mean and finite variance. 
For $k=i,i+j$, let ${\cal P}(Y_k|Y_{\condset})$ denote the best linear predictor of $Y_k$ w.r.t $Y_{\condset}$ which minimizes the mean squared error so that $\tilde{\E}_{k|{\condset}} = Y_k - {\cal P}(Y_k|Y_{\condset})$ is the corresponding prediction error. 
The partial correlation coefficient of $Y_i$ and $Y_{i+j}$ given $Y_{\condset}$ is then defined by $ 
\rho_{i,i+j; \condset} =
\corr[\tilde{\E}_{i|{\condset}},\tilde{\E}_{{i+j}|{\condset}}]$.
An equivalent definition is given as follows. 
For $i=1,\ldots,d-2$, let
\begin{align}
\label{rec1}
\begin{split}
\E_{i|i+1}  := Y_{i}-{\cal P}(Y_i|Y_{i+1}),\quad
& \text{and} \quad
\E_{i+2|i+1}  := Y_{i+2}-{\cal P}(Y_{i+2}|Y_{i+1}).
\end{split}
\end{align}
Moreover, for $j=3,\ldots,d-1$, and $i=1,\ldots,d-j$,  define
\begin{align}
\label{rec2}
\begin{split}
\E_{i|\condset} & := 
\E_{i|\sarg{i}{j-1}}- {\cal P}(\E_{i|\sarg{i}{j-1}}|\E_{i+j-1|\sarg{i}{j-1}}),
\\
\E_{i+j|\condset} & := \E_{i+j|\sarg{i+1}{j-1}}-
{\cal P}(\E_{i+j|\sarg{i+1}{j-1}}|\E_{i+1|\sarg{i+1}{j-1}}).
\end{split}
\end{align}
It is easy to show that $\E_{k|\condset}=\tilde{\E}_{k|\condset}$ for all $k=i,i+j$
  and $(i,j)\in \idxsett$.
That is, $\E_{k|\condset}$ is the error of the best linear prediction of $Y_k$ in terms of $Y_{\condset}$. 
Thus, $\rho_{i,i+j;\condset} = 
\corr[{\E}_{i|\condset},\E_{i+j|\condset}]$. 
However, the interpretation of the partial correlation coefficient as a measure of conditional dependence is different depending on whether one considers it as the correlation of  
$(\tilde{\E}_{i|\condset},\tilde\E_{i+j|\condset})$ or 
$({\E}_{i|\condset},\E_{i+j|\condset})$.
For instance,
$\rho_{14;23}  = \corr[\tilde\E_{1|23},\tilde\E_{4|23} ]$ can be interpreted as the correlation between $Y_1$ and $Y_4$ after each variable has been corrected for the linear influence of $Y_{2:3}$, i.e., 
$\corr[g(\tilde{\E}_{k|23}),h(Y_{2:3})]=0$ for all linear functions $g$ and $h$.
The idea of the partial copula is to replace the prediction errors $\E_{1|23}$ and $\E_{4|23}$ by the CPITS 
$U_{1|23}$
 and 
$U_{4|23}$
  which are  independent of $Y_{2:3}$.
On the other side, $\rho_{14;23}  = 
\corr\left[\E_{1|23},\E_{4|23} 
\right]
$ is the correlation of $(\E_{1|2},\E_{4|3})$ after $\E_{1|2}$ has been corrected for the linear influence of $\E_{3|2}$, and $\E_{4|3}$ has been corrected for the linear influence of $\E_{2|3}$. 
Consequently, a different generalization of the partial correlation coefficient emerges if we do not only decorrelate the involved random variables in \eqref{rec1} and \eqref{rec2} but render them independent by replacing each expression of the form $X-{\cal P}(X|Z)$ in \eqref{rec1} and \eqref{rec2} by the corresponding CPIT $F_{X|Z}(X|Z)$. 
The joint distribution of a resulting pair of random variables is given by the $j$-th order partial copula and the set of these copulas together with a vine structure constitute the partial vine copula.

\begin{mydef}
[Partial vine copula (\PVCA) and  $j$-th order partial copulas]
\label{def_hopartial}
Consider the  D-vine copula $C_{1:\Mydim}\in{\cal C}_d$ stated in \autoref{dvinedef}. 
In the first tree, we set for
$i=1,\ldots,d-1\colon$ $\pvc{C}_{i,i+1}= C_{i,i+1}, 
$
while in the second tree, we denote for
$i=1,\ldots,d-2, k=i,i+2\colon
\pvc{C}_{i,i+2\ps i+1}=\parsign{C}_{i,i+2\ps i+1}$ 
and
$\pvc{U}_{k|i+1}= U_{k|i+1} = F_{k|i+1}(U_k|U_{i+1}).$
In the remaining trees $j=3,\ldots,d-1,$ 
for $ i=1,\ldots,d-j$, we define 
\begin{align*}
\pvc{U}_{i|\condset}:= \pvc{F}_{i|\condset}(U_i|U_{\condset})
&:= \partial_2\pvc{C}_{i,i+j-1\ps\sarg{i}{j-1}}(\pvc{U}_{i|\sarg{i}{j-1}},
\pvc{U}_{i+j-1|\sarg{i}{j-1}}),
\\
\pvc{U}_{i+j|\condset}:= \pvc{F}_{i+j|\condset}(U_{i+j}|U_{\condset})
&:= \partial_1\pvc{C}_{i+1,i+j\ps\sarg{i+1}{j-1}}(\pvc{U}_{i+1|\sarg{i+1}{j-1}},
\pvc{U}_{i+j|\sarg{i+1}{j-1}}),
\\
\shortintertext{and}
 \pvc{C}_{i,i+j\ps\condset}(a,b) &:= 
\mathbb{P}(\pvc{U}_{i|\condset}\leq a,\pvc{U}_{i+j|\condset}\leq b).
\end{align*}
We call the resulting simplified vine copula $\pvc{C}_{1:d}$ the {partial vine copula (PVC)} of $C_{1:d}$.
Its density is given by 
\begin{align*}
\pvc{c}_{1:d}(u_{1:d}) &:= \prod_{(i,j)\in\idxset}
\pvc{c}_{i,i+j\ps \condset}(\pvc{u}_{i|\condset},
\pvc{u}_{i+j|\condset}).
\end{align*}
For $k=i,i+j$, we call $\pvc{U}_{k|\condset}$ the $(j-2)$-th order {partial probability integral transform (PPIT)} of $U_k$ w.r.t. $U_{\condset}$ and $\pvc{C}_{i,i+j\ps\condset}$
the $(j-1)$-th order {partial copula} of $F_{i,i+j|{\condset}}$ 
that is induced by $\pvc{C}_{1:d}$.
\end{mydef}

Note that the first-order partial copula coincides with the partial copula of a conditional distribution with one conditioning variable.
If $j\geq 3$, we call $\pvc{C}_{i,i+j\ps \condset}$ a higher-order partial copula.
It is easy to show that, %
for all $(i,j)\in\idxset$, $\pvc{U}_{i|\condset}$ is the CPIT of $\pvc{U}_{i|\sarg{i}{j-1}}$ w.r.t. $\pvc{U}_{i+j-1|\sarg{i}{j-1}}$ and $\pvc{U}_{i+j|\condset}$ is the CPIT of $\pvc{U}_{i+j|\sarg{i+1}{j-1}}$ w.r.t. $\pvc{U}_{i+1|\sarg{i+1}{j-1}}$.
Thus, PPITs are uniformly distributed and higher-order partial copulas are indeed copulas.
Since $\pvc{U}_{i|\condset}$ is the CPIT of 
$\pvc{U}_{i|\sarg{i}{j-1}}$
 w.r.t. $\pvc{U}_{i+j-1|\sarg{i}{j-1}}$, it is independent of $\pvc{U}_{i+j-1|\sarg{i}{j-1}}$.
However, in general it is not true that $\pvc{U}_{i|\condset}\perp U_{\condset}$ as the following proposition clarifies.
\begin{mylem}[Relation between  PPITs and  CPITs]
\label{equalppitcpit}
For $(i,j)\in \idxsett$ and $k=i,i+j,$ it holds:
\begin{align*}
\pvc{U}_{k|\condset} \indep U_{\condset}
& \eq 
\pvc{U}_{k|\condset} = U_{k|\condset}\ \text{(\as)}.
\end{align*}
\end{mylem}
\begin{showproof}
\begin{myproof}
See \appref{proof_ppitcpit}.
\end{myproof}
\end{showproof}

Note that 
$(\pvc{U}_{i|\condset},\pvc{U}_{i+j|\condset}) = (U_{i|\condset},U_{i+j|\condset})$ (a.s.) if and only if $\pvc{C}_{i,i+j\ps \condset}= \parsign{C}_{i,i+j\ps \condset}.$
 Consequently, if a higher-order partial copula does not coincide with the partial copula, it describes the distribution of a pair of uniformly distributed random variables which are neither jointly nor individually independent of the conditioning variables of the corresponding conditional copula.
Thus, if the simplifying assumption holds, then 
$C_{1:d} = \pvc{C}_{1:d}$%
, i.e., higher-order partial copulas, partial copulas and conditional copulas coincide. 
This insight is used by \citep{Kurz2017} to develop tests for the simplifying assumption in high-dimensional vine copulas.

Let $k=i,i+j$, and $\Gfunt{k}{t_k} = (\pvc{F}_{k|\condset})^{-1}(t_k|t_{\condset})$ denote the inverse of $\pvc{F}_{k|\condset}(\cdot|t_{\condset})$ w.r.t. the first argument. A $(j-1)$-th order partial copula is then given by
\begin{align*}
\begin{split}
&\pvc{C}_{i,i+j\ps\condset}(a,b) 
= \mathbb{P}(\pvc{U}_{i|\condset}\leq a, \pvc{U}_{i+j|\condset}\leq b)
= \mathbb{E}\big[\mathbb{P}(\pvc{U}_{i|\condset}\leq a, \pvc{U}_{i+j|\condset}\leq b|U_{\condset})\big]
\\
& = 
\int_{[0,1]^{j-1}}
\condcop_{i,i+j\cs\condset}
\Big(
F_{i|\condset}\big(\Gfunt{i}{a}\big|t_{\condset}\big),
F_{i+j|\condset}\big(\Gfunt{i+j}{b}\big|t_{\condset}\big)
\Big|t_{\condset}
\Big)
\text{d}F_{\condset}(t_{\condset}).
\end{split}
\end{align*}
\begin{note}
\footnote{For $j\geq 3$, the evaluation of the integrand in \eqref{comp_hopartial} requires, for $k=i,i+j$, the computation of $\pvc{G}_{k|\condset}(\cdot|t_{\condset})$ which must be obtained by inverting the $(j-2)$-dimensional integral 
$\pvc{F}_{k|\condset}(\cdot|t_{\condset})$.
Note that $\pvc{F}_{k|\condset}(\cdot|t_{\condset})$ as well is given by a $(j-2)$-dimensional integral and its integrand also requires the inversion of $(j-3)$-dimensional integrals. 
Evidently, this recursive structure of nested integrals continues if $j\geq 4$. 
Therefore, it is virtually impossible to obtain a closed-form expression for a higher-order partial copula except for very special cases.
But also the numerical approximation of the integral given in \eqref{comp_hopartial} is only feasible for very low orders since the computational complexity for the evaluation of the integrand increases tremendously with each additional order.}
\end{note}%
If $j\geq 3$, $\pvc{C}_{i,i+j\ps\condset}$ depends on 
$F_{i| \condset},F_{i+j|\condset}, \condcop_{i,i+j\cs\condset}$, and $F_{\condset}$, i.e., it depends on $C_{i:i+j}$.
Moreover, $\pvc{C}_{i,i+j\ps\condset}$ also depends on $\pvc{G}_{i|\condset}$ and $\pvc{G}_{i+j|\condset}$, which are determined by the regular vine structure.
Thus, the corresponding PVCs of different regular vines may be different.
In particular, if the \SA{} does not hold, higher-order partial copulas of different PVCs which refer to the same conditional distribution may not be identical. 
This is different from the partial correlation coefficient or the partial copula which do not depend on the structure of the regular vine.

In general, higher-order partial copulas do not share the simple interpretation of the partial copula because they can not be considered as expected conditional copulas. However, higher-order partial copulas can be more attractive from a practical point of view. 
The estimation  of the partial copula of $\condcop_{i,i+j\cs \condset}$ requires the estimation of the two $j$-dimensional conditional cdfs $F_{i|\condset}$ and $F_{i+j|\condset}$ to construct pseudo-observations from the CPITs $(U_{i|\condset},U_{i+j|\condset})$.
As a result, a non-parametric estimation of the partial copula is only sensible if $j$ is very small. 
In contrast, a higher-order partial copula is the distribution of two PPITs  $(\pvc{U}_{i|\condset},\pvc{U}_{i+j|\condset})$ which are made up of only two-dimensional functions (\autoref{def_hopartial}).
Thus, the non-parametric estimation of a higher-order partial copula does not suffer from the curse of dimensionality and is also sensible for large $j$ \citep{Nagler2015}.
But also in a parametric framework the specification of the model family is much easier for a higher-order partial copula than for a conditional copula.
This renders higher-order partial copulas very attractive from a modeling point of view to analyze and estimate bivariate conditional dependencies.
As we show in \autoref{practice}, the \PVCA{} is also the probability limit of many estimators of pair-copula constructions and thus of great practical importance.

\section{Properties of the \pvca{} and examples}
\label{hop_pvc_propex}
In this section, we analyze to what extent the \PVCA{} describes the dependence structure of the data generating copula if the \SA{} does not hold.
We first investigate whether the bivariate margins of $\pvc{C}_{1:d}$ match the bivariate margins of $C_{1:d}$ and then take a closer look at conditional independence relations.
By construction, the bivariate margins $\pvc{C}_{i,i+1}, i=1,\ldots,d-1,$ of the \PVCA{} given in \autoref{def_hopartial} are identical to the corresponding margins $C_{i,i+1}, i=1,\ldots,d-1,$ of $C_{1:d}$.
That is because the \PVCA{} explicitly specifies these $d-1$ margins in the first tree of the vine. 
The other bivariate margins $\pvc{C}_{i,i+j}$, where $(i,j)\in \idxsett$, are implicitly specified and  given by
\begin{align*}
\pvc{C}_{i,i+j}(u_i,u_{i+j}) & = \int_{[0,1]^{j-1}} \pvc{C}_{i,i+j\ps \condset}(\pvc{F}_{i|\condset}(u_i|u_{\condset}),
\pvc{F}_{i+j}(u_{i+j}|u_{\condset}))\text{d}\pvc{C}_{\condset}(u_{\condset}).
\end{align*}
The relation between the implicitly given bivariate margins of the \PVCA{} and the underlying copula are summarized in the following lemma. 
\begin{mylem}[Implicitly specified margins of the \PVCA{}]
\label{propmarg}
Let  $C_{1:d}\in{\cal C}_d\backslash \SPCC{C}_d$, 
$(i,j)\in \idxsett$,
and $\tau_E$ and $\rho_E$ denote Kendall's $\tau$ and Spearman's $\rho$ of the copula $E\in {\cal C}_2$.
In general, it holds that $\pvc{C}_{i,i+j} \neq C_{i,i+j}, \rho_{\pvc{C}_{i,i+j}}\neq \rho_{C_{i,i+j}}$, and $\tau_{\pvc{C}_{i,i+j}}\neq \tau_{C_{i,i+j}}$.
\end{mylem}

\begin{showproof}
\begin{myproof}
See \appref{prd3}.
\end{myproof}
\end{showproof}

The next example provides an example of a three-dimensional \PVCA{} and illustrates the results of \autoref{propmarg}. 
{Other examples of \PVCA{s} in three dimensions are given in \citet{Spanhel2015c}.}
\begin{myex}
\label{imp_margin}
Let $C^{FGM_{2}}(\theta)$ denote the bivariate FGM copula 
\begin{align*}
C^{FGM_{2}}(u_1,u_2;\theta) & = u_1u_2[1+\theta(1-u_1)(1-u_2)], \quad |\theta|\leq 1,
\end{align*}
and $C^{A}(\gamma)$ denote the following asymmetric version of the FGM copula (\cite{Nelsen2006}, Example 3.16)
\begin{align}
\label{asyfgm}
C^{A}(u_1,u_2;\gamma) & = u_1u_2[1+ \gamma u_1(1-u_1)(1-u_2)], \quad |\gamma|\leq 1.
\end{align}
Assume that $C_{\idxz\idxo} = C^{A}({\gamma}), C_{\idxo\idxt} = C^\perp,  \condcop_{\idxz\idxt\cs \idxo}(\cdot,\cdot\cs u_2) = C^{FGM_{2}}{(\cdot,\cdot ; 1-2u_\idxo)}$ for all $u_2$, so that
\begin{align*}
C_{1:3}(u_{1:3}) & = \int_{0}^{u_2} C^{FGM_{2}}{(\partial_2C^{A}(u_1,t_2),u_3 ; 1-2t_\idxo)}\text{d}t_2. 
\end{align*}
Elementary computations show that the implicit margin is given by
\begin{align*}
C_{\idxz\idxt}(u_\idxz,u_\idxt) & = u_\idxz u_\idxt[
\gamma(u_1-3u_1^2+2u_1^3)(1-u_3) +3]/3,
\end{align*}
which is a copula with quartic sections in $u_\idxz$ and square sections in $u_\idxt$ if $\gamma\neq 0$. 
The corresponding \PVCA{} is
\begin{align*}
\pvc{C}_{1:3}(u_{1:3}) & = \int_0^{u_2} \pvc{C}_{13\ps 2}
(F_{1|2}(u_1|t_2),F_{3|2}(u_3|t_2))\text{d}t_2 \stackrel{\pvc{C}_{13\ps 2}=C^{\indep}}{=}
u_3\int_{0}^{u_2} \partial_2C^{A}(u_1,t_2)\text{d}t_2
\end{align*}
and the implicit margin of $\pvc{C}_{1:3}$ is
\begin{align*}
\pvc{C}_{\idxz\idxt}(u_\idxz,u_\idxt) & = \pvc{C}_{1:3}(u_{1},1,u_3) = 
u_\idxz u_\idxt.
\end{align*}
Moreover, $
\rho_{{C}_{13}} = -\gamma/1080,  \tau_{C_{13}} = -\gamma/135, 
$
but $\rho_{\pvc{C}_{13}} = \tau_{\pvc{C}_{13}} = 0$.
\end{myex}

Higher-order partial copulas can also be used to construct new measures of conditional dependence. 
For instance, if $X_{1:d}$ is a random vector with copula $C_{1:d}\in{\cal C}_d$,
higher-order partial Spearman's $\rho$  and Kendall's $\tau$ of $X_i$ and $X_{i+j}$ given $X_{\condset}$ are defined by
\begin{align*}
{\tau}_{\pvc{C}_{i,i+j\ps\condset}} & = 4\int_{[0,1]^2} \pvc{C}_{i,i+j\ps \condset}(a,b)
\text{d}\pvc{C}_{i,i+j\ps \condset}(a,b)-1,
\\
{\rho}_{\pvc{C}_{i,i+j\ps \condset}}  
& = 12 \int_{[0,1]^2}  \pvc{C}_{i,i+j\ps \condset}(a,b)\text{d}a\text{d}b-3.
\end{align*}
Note that all dependence measures that are derived from a higher-order partial copula are defined w.r.t. a regular vine structure and that they coincide with their conditional analogues if the \SA{} holds.
A partial correlation coefficient of zero is commonly interpreted as an indication of conditional independence, although this can be quite misleading if the underlying distribution is not close to a Normal distribution (\citet{Spanhel2015c}). Therefore, one might wonder to what extent higher-order partial copulas can be used to check for conditional independencies.  
If $\pvc{C}_{i,i+j\ps \condset}$ equals the independence copula, we say that $X_{i}$ and $X_{i+j}$ are ($j$-th order) partially independent given $X_{\condset}$ and write
$X_i \stackrel{\text{\tiny PVC}}{\indep} X_{i+j}|X_{\condset}$.
The following theorem establishes that there is in general no relation between conditional independence and higher-order partial independence.%
\begin{mythe}
[{Conditional independence and {$j$-th order partial independence}}]
\label{cond_indep}
Let $d\geq 4$, $(i,j)\in\idxset$, and $C_{1:\Mydim}\in{\cal C}_d\backslash \SPCC{C}_d$ be the copula of $X_{1:d}$. 
It holds that
\begin{align*}
X_i \indep X_{i+2}|X_{i+1} &\imp X_i \stackrel{\text{\tiny{PVC}}}{\indep} X_{i+2}|X_{i+1},
\\ \forall j\geq 3: X_i \indep X_{i+j}|X_{\condset}  &\not\Rightarrow 
X_i \stackrel{\text{\tiny{PVC}}}{\indep} X_{i+j}|X_{\condset},
\shortintertext{and}
\forall j\geq 2: X_i \indep X_{i+j}|X_{\condset}  &\not\Leftarrow  
X_i \stackrel{\text{\tiny{PVC}}}{\indep} X_{i+j}|X_{\condset}.
\end{align*}
\end{mythe}
\begin{showproof}
\begin{myproof}
See \appref{Proof_cond_indep}.
\end{myproof}
\end{showproof}

The next five-dimensional example illustrates higher-order partial copulas, higher-order PPITs, and the relation between partial independence and conditional independence.

\begin{myex}
\label{fgm_ex5}
Consider the following exchangeable D-vine copula $C_{1:5}$ which does not satisfy the simplifying assumption:
\begin{align}
C_{12}=C_{23} = C_{34} = C_{45}, \ \
\condcop_{13\cs 2}\ \!\!&= \ \!\!\condcop_{24\cs 3} = \condcop_{35\cs 4},\ \
\condcop_{14\cs 2:3} = \condcop_{25\cs 3:4}, \notag
\\
C_{12}\ &  =\ C^\perp, \label{eqCPIT}
\\
\condcop_{13\cs 2}(a,b|u_{2})\ & =\ C^{FGM_{2}}(a,b\ \! ;1-2u_2), 
\quad \forall (a,b,u_2)\in[0,1]^3
\label{uncondfgm}
\\
\condcop_{14\cs 2:3}\ & =\ C^\perp,
\label{uncondfgm2}
\\
\condcop_{15\cs 2:4}\ & =\ C^\perp,
\label{uncondfgm3}
\end{align}
where $\condcop_{i,i+j\cs\condset}=C^{\indep}$ means that
$\condcop_{i,i+j\cs\condset}(a,b|u_{\condset})=ab$ for all 
$(a,b,u_{\condset})\in[0,1]^{j+1}$.
\end{myex}

\begin{figure}
\newcommand{\mys}{0.5}
\begin{subfigure}[b]{\mys\textwidth}
\centering
  \includegraphics{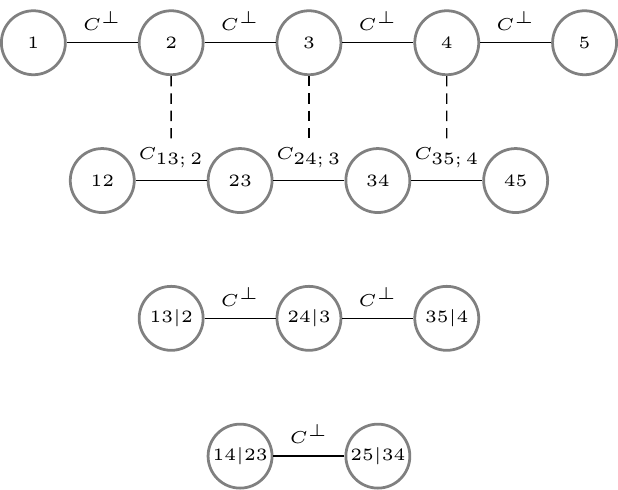}
  \label{fig_5fgm1.pdf}
  \caption{Vine copula in \autoref{fgm_ex5}.}
  \end{subfigure}%
\begin{subfigure}[b]{\mys\textwidth}
  \centering
  \includegraphics{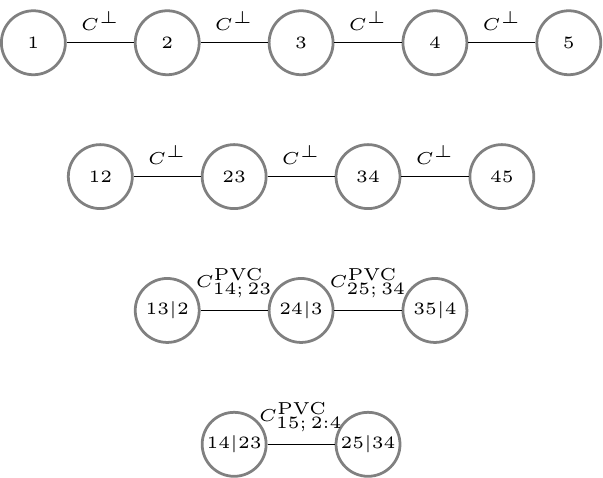}
  \label{fig_5fgm2.pdf}
  \caption{\PVCA{} of \autoref{fgm_ex5}.}
\end{subfigure}

\caption{The non-simplified D-vine copula given in \autoref{fgm_ex5} and its \PVCA{}. The influence of conditioning variables on the conditional copulas is indicated by dashed lines.}\label{fig_ex_3_3}
\end{figure}

All conditional copulas of the vine copula in \autoref{fgm_ex5} correspond to the independence copula except for the second tree. 
Note that for all $i=1,2,3$, $(U_{i},U_{i+1},U_{i+2})\sim C^{FGM_{3}}(1)$, where
$C^{FGM_{3}}(u_{1:3};\theta) = \prod_{i=1}^3u_i+\theta\prod_{i=1}^3u_i(1-u_i), |\theta|\leq 1$, is the three-dimensional FGM copula.
The left panel of \autoref{fig_ex_3_3} illustrates the D-vine copula of the data generating process. 
We now investigate the \PVCA{} of $C_{1:5}$ which is illustrated in the right panel of \autoref{fig_ex_3_3}.
{Since $C_{1:5}$ and $\pvc{C}_{1:5}$ are exchangeable copulas, we only report the PPITs $\pvc{U}_{1|2},\pvc{U}_{1|2:3}$ and $\pvc{U}_{1|2:4}$ in the following lemma.}

\begin{mylem}[The \PVCA{} of \autoref{fgm_ex5}]
\label{spurious_dep}
Let $C_{1:5}$ be defined as in \autoref{fgm_ex5}. Then
\begin{align*}
\pvc{C}_{12}=\pvc{C}_{23} = \pvc{C}_{34} = \pvc{C}_{45}, \ \ 
\pvc{C}_{13\ps 2}\ \!\! & =\ \!\!\pvc{C}_{24\ps 3} = \pvc{C}_{35\ps 4},\ \ 
\pvc{C}_{14\ps 2:3} = \pvc{C}_{25\ps 3:4},
\\
\pvc{C}_{12}\ & =\ C^\perp,
\\
\pvc{C}_{13\ps 2}\ & =\ C^{\perp},
\\
\pvc{C}_{14\ps 2:3}(a,b)\ & =\ 
 C^{FGM_{2}}(a,b\ \!;1/9),
 \quad \forall (a,b)\in[0,1]^2
\\
\pvc{C}_{15\ps 2:4}\ & \neq\ C^\perp,
\shortintertext{and}
\pvc{U}_{1|2}  = U_1 \ &=\  U_{1|2},
\\
\pvc{U}_{1|2:3}  = U_1\ &\neq \ U_{1|2:3} = U_{1}[1+(1-U_{1})
(1-2U_{2})(1-2U_{3})],
\\
\pvc{U}_{1|2:4}  = U_{1}[1+\theta(1-U_{1})(1-2U_{4})]\ &\neq \ U_{1|2:4} = U_{1|2:3}.
\end{align*}
\end{mylem}
\begin{showproof}
\begin{myproof}
See \appref{Derive_spurious_dep}.
\end{myproof}
\end{showproof}

\autoref{spurious_dep}  demonstrates that $j$-th order partial copulas may not be independence copulas, although the corresponding conditional copulas are independence copulas.
In particular, under the data generating process the edges of the third tree of $C_{1:5}$ are independence copulas. 
Neglecting the conditional copulas in the second tree and replacing them with first-order partial copulas induces spurious dependencies in the third tree of $\pvc{C}_{1:5}$.
The introduced spurious dependence also carries over to the fourth tree where we have (conditional) independence in fact.
{
Nevertheless, the PVC reproduces the bivariate margins of $C_{1:5}$ pretty well.
It can be readily verified that $(\pvc{C}_{13},\pvc{C}_{14},\pvc{C}_{24},\pvc{C}_{25},\pvc{C}_{35})
= (C_{13},C_{14},C_{24},C_{25},C_{35})$, i.e., except for $\pvc{C}_{15}$, all bivariate margins of $\pvc{C}_{1:5}$  match the bivariate margins of $C_{1:5}$ in \autoref{fgm_ex5}. 
Moreover,  the mutual information in the third and fourth tree are larger if higher-order partial copulas are used instead of the true conditional copulas. Thus, the spurious dependence in the third and fourth tree decreases the Kullback-Leibler divergence from $C_{1:5}$ and therefore acts as a countermeasure for the spurious (conditional) independence in the second tree.
}
\autoref{spurious_dep} also reveals that $U_{1|2:4}$ is a function of $U_{2}$ and $U_3$, i.e. the true conditional distribution function $F_{1|2:4}$ depends on $u_2$ and $u_3$. In contrast, $\pvc{F}_{1|2:4}$, the resulting model for $F_{1|2:4}$ which is implied by the \PVCA{}, depends only on $u_4$.
That is, the implied conditional distribution function of the \PVCA{} depends on the conditioning variable which actually has no effect.

\section{Approximations based on the \pvca{}}
\label{sec_simp_approx_3d}%

The specification and estimation of \SVCM{s} is commonly based on procedures that asymptotically minimize the Kullback-Leibler divergence (KLD) in a stepwise fashion. For instance, if a parametric vine copula model is used, the step-by-step ML {estimator} (\citet{HobkHaff2012,HobkHaff2013}), where one estimates tree after tree and sequentially minimizes the estimated KLD conditional on the estimates {from} the previous trees, is often employed in order to select and estimate the parametric pair-copula families of the vine.  
But also the non-parametric methods of \citet{Kauermann2013b} and \citet{Nagler2015} proceed in a stepwise manner and asymptotically minimize the KLD of each pair-copula separately under appropriate conditions.
{
In this section, we investigate the role of the \PVCA{} when it comes to approximating non-simplified vine copulas.} 

Let $C_{1:d}\in {\cal C}_d$ and $ \svcs{C}_{1:d}\in\SPCC{C}_d$.
The KLD of $\svcs{C}_{1:d}$ from the true copula $C_{1:d}$ is given by
\begin{align*}
\KL{C_{1:d}}{\svcs{C}_{1:d}} & 
= \mathbb{E}\left[\log \frac{c_{1:d}(U_{1:d})}{\svcs{c}_{1:d}(U_{1:d})}
\right],
\end{align*}
where the expectation is taken w.r.t. the true distribution $C_{1:d}$.
We now decompose the KLD into the Kullback-Leibler divergences related to each of the $d-1$ trees. 
For this purpose, let $j=1,\ldots,d-1$ and define
\begin{align*}
\allvines{j}&:=\left
\{(\svcs{C}_{i,i+j\ps \condset})_{i=1,\ldots,d-j}\colon 
\svcs{C}_{i,i+j\ps \condset}\in {\cal C}_{2} \text{ for } 1\leq i\leq d-j
\right
\},
\end{align*}
so that 
$\treeseq{j} = \times_{k=1}^{j}\allvines{k}$ represents all possible \SVCM{s}
up to and including the $j$-th tree. 
Let ${{\cal T}}_{j}\in\allvines{j}, 
{{\cal T}}_{1:j-1}\in\treeseq{j-1}$. 
The KLD of the \SVCM{} associated with ${{\cal T}}_{1:d-1}$ is given by
\begin{align}
\KL{C_{1:d}}{{{\cal T}}_{1:d-1}}
&=\sum_{j=1}^{d-1}D_{KL}^{(j)}({{\tree{j}}} ({{\cal T}}_{1:j-1})),
\label{sbs3}
\shortintertext{where}
D_{KL}^{(1)}({{\cal T}}_1({{\cal T}}_{1:0}))) & := 
D_{KL}^{(1)}({\cal T}_1):=\sum_{i=1}^{d-1}
\mathbb{E}
\left[
\log \frac{c_{i,i+1}(U_i,U_{i+1})}{\svcs{c}_{i,i+1}(U_i,U_{i+1})}
\right]\notag
\end{align}
denotes the KLD related to the first tree,
and for the remaining trees $j=2,\ldots,d-1$, the related KLD is
\begin{align*}
D_{KL}^{(j)}({\tree{j}} ({{\cal T}}_{1:j-1})) 
& := \sum_{i=1}^{d-j} \mathbb{E}
\left[\log \frac
{
\condden_{i,i+j\cs \condset}(U_{i|\condset},
U_{i+j|\condset}|U_{\condset})
}
{
\svcs{c}_{i,i+j\ps \condset}(\svcs{U}_{i|\condset},
\svcs{U}_{i+j|\condset})
}
\right].
\end{align*}
For instance, if $d=3$, the KLD can be decomposed into the KLD related to the first tree $D_{KL}^{(1)}$ and  to the second tree $D_{KL}^{(2)}$ as follows
\begin{align*} 
\KL{C_{\idxlist}}{{\cal T}_{1:2}}
& = \KL{C_{\idxlist}}{({\cal T}_{1},{\cal T}_2)}
=  D_{KL}^{(1)}({{\cal T}}_1)
+ D_{KL}^{(2)}({{\cal T}}_2({{\cal T}}_{1}))
\\
&=
\mathbb{E}\left[\log \frac{{c}_{\idxz\idxo}(U_{\idxz:\idxo}){c}_{\idxo\idxt}(U_{\idxo:\idxt})}
{\svcs{c}_{\idxz\idxo}(U_{\idxz:\idxo})\svcs{c}_{\idxo\idxt}(U_{\idxo:\idxt})}\right]
+ \mathbb{E}\left[\log \frac{{\condden}_{\idxz\idxt\cs \idxo}
\big(\partial_2{C}_{\idxz\idxo}(U_{\idxz:\idxo}),
\partial_1{C}_{\idxo\idxt}(U_{\idxo:\idxt})|U_\idxo
\big)}{\svcs{c}_{\idxz\idxt\ps \idxo}
\big(\partial_2\svcs{C}_{\idxz\idxo}(U_{\idxz:\idxo}),
\partial_1\svcs{C}_{\idxo\idxt}(U_{\idxo:\idxt})
\big)}\right].
\end{align*}
Note that the KLD related to tree $j$ 
depends on the specified copulas in the lower trees because they determine at which values the copulas in tree $j$ are evaluated.
{The following theorem shows that,
if one sequentially minimizes the KLD related to each tree, then the optimal \SVCM{} is the PVC.
}
\begin{mythe}
[Tree-by-tree KLD minimization using the PVC]
\label{hop}
Let 
$C_{1:d}\in {\cal C}_\Mydim$ be the data generating copula and
$\pvc{{\cal T}}_j
:=(\pvc{C}_{i,i+j\ps \condset})_{i=1,\ldots,d-j}$,  so that 
$\mpartreeseq{j}  := 
\times_{k=1}^{j}\pvc{{\cal T}}_{k}$
collects all copulas of the PVC up to and including the $j$-th tree. 
It holds that
\begin{align}
\forall j=1,\ldots,d-1
\colon\quad
\std{\arg\min}{\tree{j}\in\allvines{j}}\quad\!
D_{KL}^{(j)}(\tree{j} (\mpartreeseq{j-1})) 
&= \pvc{{\cal T}}_{j}.  \label{sbs1}
\end{align}
\end{mythe}
\begin{showproof}
\begin{myproof}
See \appref{proof_seqkl}.
\end{myproof}
\end{showproof}

According to \autoref{hop}, if the true copulas are specified in the first tree, one should choose the first-order partial copulas in the second tree, the second-order partial copulas in the third tree etc. to minimize the KLD tree-by-tree.
\autoref{hop} also remains true if we replace ${\cal C}_2$ in the definition of 
$\allvines{j}$ by the space of absolutely continuous bivariate cdfs. 
The \PVCA{}  ensures that random variables in higher trees are uniformly distributed since the resulting random variables in higher trees are higher-order PPITs.
If one uses a different approximation, such as the one used by \citet{HobkHaff2010} and \citet{Stober2013b}, then the random variables in higher trees are not necessarily uniformly distributed and pseudo-copulas (\citet{Fermanian2012}) can be used to further minimize the KLD. 
\citet{Stober2013b} note in their appendix that if $C_{\idxlist}$ is a FGM copula and the copulas in the first tree are correctly specified, then the KLD from the true distribution has an extremum at 
$\svcs{C}_{\idxz\idxt\ps \idxo} = C^\perp = \pvc{C}_{\idxz\idxt\ps \idxo}$.
{
If $\condcop_{13\cs 2}$ belongs to a parametric family of bivariate copulas whose  parameter depends on $u_2$, then $\pvc{C}_{13\ps 2}$ is in general not a member of the same copula family with a constant parameter, see \citet{Spanhel2015c}.
Together with \autoref{hop} it follows that the proposed \svca{s} of \citet{HobkHaff2010} and \citet{Stober2013b} can be improved if the first-order partial copula is chosen in the second tree, and not a copula of the same parametric family as the conditional copula but with a constant dependence parameter such that the KLD is minimized. 

Besides its interpretation as generalization of the partial correlation matrix, the \PVCA{} can also be interpreted as the \SVCM{} that minimizes the KLD tree-by-tree.
This sequential minimization neglects that the KLD related to a tree depends on the copulas that are specified in the former trees. 
For instance, if $d=3$, the KLD of the first tree
$
D_{KL}^{(1)}({\cal T}_1)
$  is minimized over the copulas 
$({\svcs{C}_{\idxz\idxo},\svcs{C}_{\idxo\idxt}})$
in the first tree ${\cal T}_1$, but the effect of the chosen copulas in the first tree ${\cal T}_1$ on the KLD related to the second tree
$
 D_{KL}^{(2)}(
{{{\cal T}}_2({{\cal T}}_{1}))}
$ is not taken into account.
Therefore, we now analyze whether the \PVCA{} also globally minimizes the KLD.
Note that specifying the wrong margins in the first tree ${\cal T}_1$, e.g.,
$
(\svcs{C}_{\idxz\idxo},\svcs{C}_{\idxo\idxt})
\neq (C_{\idxz\idxo},C_{\idxo\idxt})
$,
increases 
$D_{KL}^{(1)}({\cal T}_1)$ in any case.
Thus, without any further investigation, it is absolutely indeterminate whether the definite increase in 
$D_{KL}^{(1)}({\cal T}_1)$ 
can be overcompensated by a possible decrease in 
$D_{KL}^{(2)}({\cal T}_2({\cal T}_1))$
if another approximation is chosen. 
The next theorem shows that the \PVCA{} is in general not the global minimizer of the KLD.

\begin{mythe}
[Global  KLD minimization if $C_{1:d}\in \SPCC{C}_d$
or $C_{1:d}\in {\cal C}_d\backslash \SPCC{C}_d$]
\label{prop_kl}
If $C_{1:d}\in\SPCC{C}_d$, i.e.,  the simplifying assumption holds for $C_{1:d}$, then
\begin{align}
\label{gkl1}
\std{\arg\min}
{\svcs{C}_{1:d}\in \SPCC{C}_d} 
\KL{C_{1:d}}
{\svcs{C}_{1:d}} & = 
\pvc{C}_{1:d}.
\end{align}
If the simplifying assumption does not hold for $C_{1:d}$, then 
$\pvc{C}_{1:d}$ might not be a global minimum. That is, $\exists C_{1:d}\in{\cal C}_d \backslash\SPCC{C}_d$ such that
\begin{align}
\label{gkl2}
\std{\arg\min}
{\svcs{C}_{1:d}\in \SPCC{C}_d} 
\KL{C_{1:d}}
{\svcs{C}_{1:d}} & \neq 
\pvc{C}_{1:d},
\end{align}
and $\forall ({\cal T}_2,\ldots,{\cal T}_{d-1})\in
\times_{k=2}^{d-1}\allvines{k} $
\begin{align}
\label{gkl3}
\std{\arg\min}
{\approxsign{\cal T}_{1:d-1} \in \treeseq{d-1} } 
\KL{C_{1:d}}
{\approxsign{{\cal T}}_{1:d-1}} & \neq 
(\pvc{{\cal T}}_1,{\cal T}_2,\ldots,{\cal T}_{d-1}).
\end{align}
\end{mythe}

\begin{showproof}
\begin{myproof}
See \appref{proof_prop_kl}. 
\end{myproof}
\end{showproof}

\autoref{prop_kl} states that, if the simplifying assumption does not hold, the KLD may not be minimized by choosing the true copulas in the first tree, first-order partial copulas in the second tree and higher-order partial copulas in the remaining trees (see \eqref{gkl2}).
It follows that, if the objective is the minimization of the KLD, it may not be optimal to specify the true copulas in the first tree, no matter what bivariate copulas are specified in the other trees (see \eqref{gkl3}).
This rather puzzling result can be explained by the fact that, if the simplifying assumption does not hold, then the approximation error of the implicitly modeled bivariate margins 
is not minimized (see \autoref{propmarg}).
For instance, if $d=3$, a departure from the true copulas $(C_{12},C_{23})$ in the first tree increases the KLD related to the first tree, but it can decrease the KLD of the implicitly modeled margin
$\svcs{C}_{\idxz\idxt}$ from $C_{\idxz\idxt}$. 
As a result, the increase in $D_{KL}^{(1)}$ can be overcompensated by a larger decrease in $D_{KL}^{(2)}$, so that the KLD can be decreased.

\autoref{prop_kl} does not imply that the \PVCA{} never minimizes the KLD from the true copula. 
For instance, if $d=3$ and if $\pvc{C}_{\idxz\idxt\ps \idxo} = C^\perp$, then 
$\KL{C_{\idxlist}}{\pvc{C}_{1:3}}$ is an extremum, which directly follows from equation \eqref{sbs1} since
\begin{align*}
\std{\arg\min}{{\cal T}_1\in \allvines{1}} \
\KL{C_{\idxlist}}
{({\cal T}_1,(C^\perp))}
& = \ \
\std{\arg\min}{{\cal T}_1\in \allvines{1}} \
D_{KL}^{(1)}({{{{\cal T}}_1}}).
\end{align*}

It is an open problem whether and when the \PVCA{} can be the global minimizer of the KLD.
Unfortunately, the \svca{} that globally minimizes the KLD is not tractable.
{However, if the \svca{} that minimizes the KLD does not specify the true copulas in the first tree, the random variables in the higher tree are not CPITs.
Thus, it is not guaranteed that these random variables are uniformly distributed and we could further decrease the KLD by assigning pseudo-copulas
(\citet{Fermanian2012}) to the edges in the higher trees.}
It can be easily shown that the resulting best approximation is then a pseudo-copula.
Consequently, the  best approximation satisfying the simplifying assumption is in general not an \SVCM{} but a simplified vine pseudo-copula if one considers the space of regular vines where each edge corresponds to a bivariate cdf.

While the \PVCA{} may not be the best approximation in the space of \SVCM{s}, it is the best feasible \SVCM{} approximation in practical applications.
That is because the stepwise specification and estimation of an \SVCM{} is also feasible for (very) large dimensions which is not true for a joint specification and estimation.
For instance, if all pair-copula families of a parametric vine copula are chosen simultaneously and the selection is done by means of information criteria,  we have to estimate $K^{d(d-1)/2}$ different models, where $d$ is the dimension and $K$ the number of possible pair-copula families that can be assigned to each edge.
On the contrary, a stepwise procedure only requires the estimation of $Kd(d-1)/2$ models.
To illustrate the computational burden, consider the \texttt{R}-package {\tt VineCopula} \citep{vinecopula} where $K=40$. For this number of pair-copula families, a joint specification requires the estimation of 64,000 ($d=3$) or more than four billion ($d=4$) models whereas only 120 ($d=3$) or 240 ($d=4$) models are needed for a stepwise specification.
For many non-parametric estimation approaches (kernels \citep{Nagler2015}, empirical distributions \citep{HobkHaff2015}), only the sequential estimation of an \SVCM{} is possible.
The only exception is the spline-based approach of  \citet{Kauermann2013b}. 
However, due to the large number of parameters and the resulting computational burden, a joint estimation is  only feasible for $d\leq 5$ \citep{Kauermann2013c}.

\section{Convergence to the \pvca{}}
\label{practice}

If the \dgp{} satisfies the \SA{}, consistent stepwise procedures for the specification and estimation of parametric and non-parametric simplified vine copula models asymptotically minimize the KLD from the true copula. 
\autoref{hop} implies that this is not true in general if the \dgp{} does not satisfy the \SA{}. 
An implication of this result for the application of \SVCM{s} is pointed out in the next corollary.

\begin{mycor}
\label{non_consistency}
{
Denote the sample size by $N$.
Let $C_{1:d}\in {\cal C}_{d}$
 be the data generating copula and $\svcs{C}_{1:d}(\theta)\in \SPCC{C}_{d}, \theta\in \Theta$, be a parametric \SVCM{}
so that 
$\exists_1 \pvcl{\theta} \in\Theta:\svcs{C}_{1:d}(\pvcl{\theta}) = \pvc{C}_{1:d}$.
The pseudo-true parameters which minimize the KLD from the true distribution are assumed to exist (see \citet{White1982} for sufficient conditions) and denoted by
\begin{align*}
\theta^{\star} & =
\ \ \std{\arg\min}{\theta\in\Theta}\ \ \KL{C_{1:d}}{\svcs{C}_{1:d}(\theta)}.
\end{align*}
Let $\hat{\theta}_S$ denote the (semi-parametric) step-by-step ML estimator and $\hat{\theta}_J$ denote the (semi-parametric) joint ML estimator defined in \citet{HobkHaff2012,HobkHaff2013}.
Under regularity conditions (e.g., Condition 1 and Condition 2 in \citep{Spanhel2016b}) and for $N\to\infty$, it holds that:
\begin{enumerate}[(i)]
\item \label{nonconnew} 
$\hat{\theta}^{S}\stackrel{p}{\to}\pvc{\theta}.$
\item \label{whitelim}
$\hat{\theta}^{J}\stackrel{p}{\to}{\theta}^\star.$
\item 
 $\exists C_{1:d}\in{\cal C}_d \backslash\SPCC{C}_d$ such that
 $\hat{\theta}^S
\not\stackrel{p}{\to} 
\theta^{\star}.$
\end{enumerate}
}
\end{mycor}

\begin{showproof}
\begin{myproof}
See \appref{proof_convergence2pvc}.
\end{myproof}
\end{showproof}

\autoref{non_consistency} shows that the step-by-step and joint ML estimator may not converge to the same limit (in probability) if the \SA{} does not hold for the data generating vine copula.
For this reason, we investigate in the following the difference between the step-by-step and joint ML estimator in finite samples. 
Note that the convergence of kernel-density estimators to the \PVCA{} has been recently established by \citet{Nagler2015}.
However, in this case, only a sequential estimation of a simplified vine copula is possible and thus the best feasible approximation in the space of simplified vine copulas is given by the \PVCA{}.

\subsection{Difference between step-by-step and joint ML estimates}
We compare the step-by-step and the joint ML estimator under the assumption that the pair-copula families of the PVC are specified for the parametric vine copula model. 
For this purpose, we simulate data from two three-dimensional copulas $C_{1:3}$ with sample sizes $N =\ 500, 2500, 25000$, perform a step-by-step and joint ML estimation, and repeat this 1000 times. 
For ease of exposition and because the qualitative results are not different, we consider copulas where $C_{12} = C_{23}$ and only
present the estimates for $(\theta_{12},\theta_{13;2})$.

\begin{myex}[\PVCA{} of the Frank copula]
\label{exA}
Let $C^{\text{Fr}}(\theta)$ denote the bivariate Frank copula with dependence parameter $\theta$ and ${C}^{\text{P-Fr}}(\theta)$ be the partial Frank copula  \citep{Spanhel2015c} with dependence parameter $\theta$. 
Let 
$C_{1:3}$ be the true copula with $
(C_{12},C_{23},C_{13\ps 2}) 
= (C^{\text{Fr}}(5.74),C^{\text{Fr}}(5.74),{C}^{\text{P-Fr}}( 5.74))$, i.e., $C_{1:3}=\pvc{C}_{1:3}$, 
  and 
$\svcs{C}_{1:3}(\theta) = 
(C^{\text{Fr}}(\theta_{12}),C^{\text{Fr}}(\theta_{23}),{C}^{\text{P-Fr}}( \theta_{13;2}))$ be the parametric \SVCM{} that is fitted to data generated from $C_{1:3}$.
\end{myex}
\begin{figure}[h!]
\centering

\begin{subfigure}[t]{0.5\textwidth}
  \centering
    \includegraphics[scale=\scaleit]{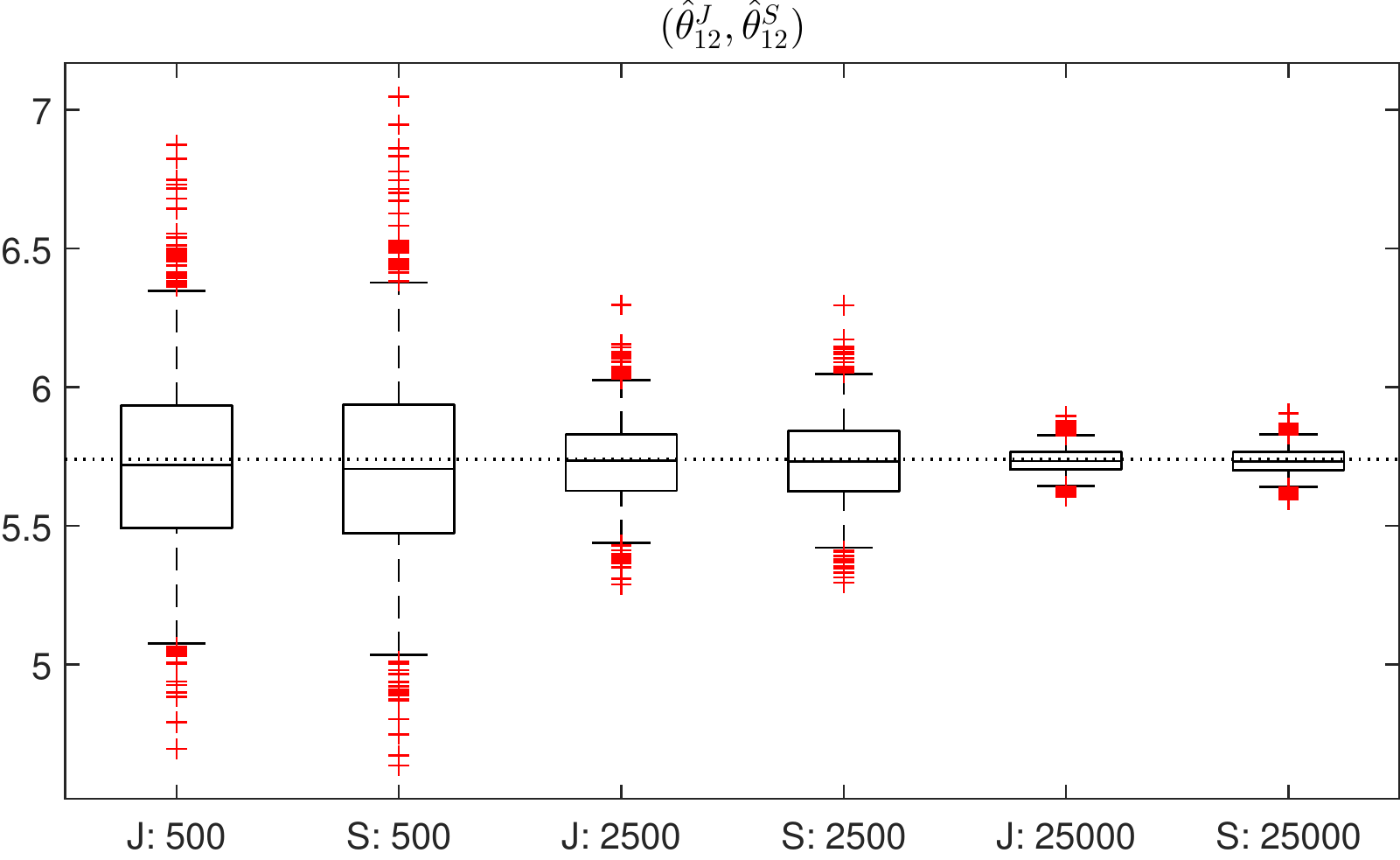}
\end{subfigure}%
\begin{subfigure}[t]{0.5\textwidth}
  \centering
    \includegraphics[scale=\scaleit]{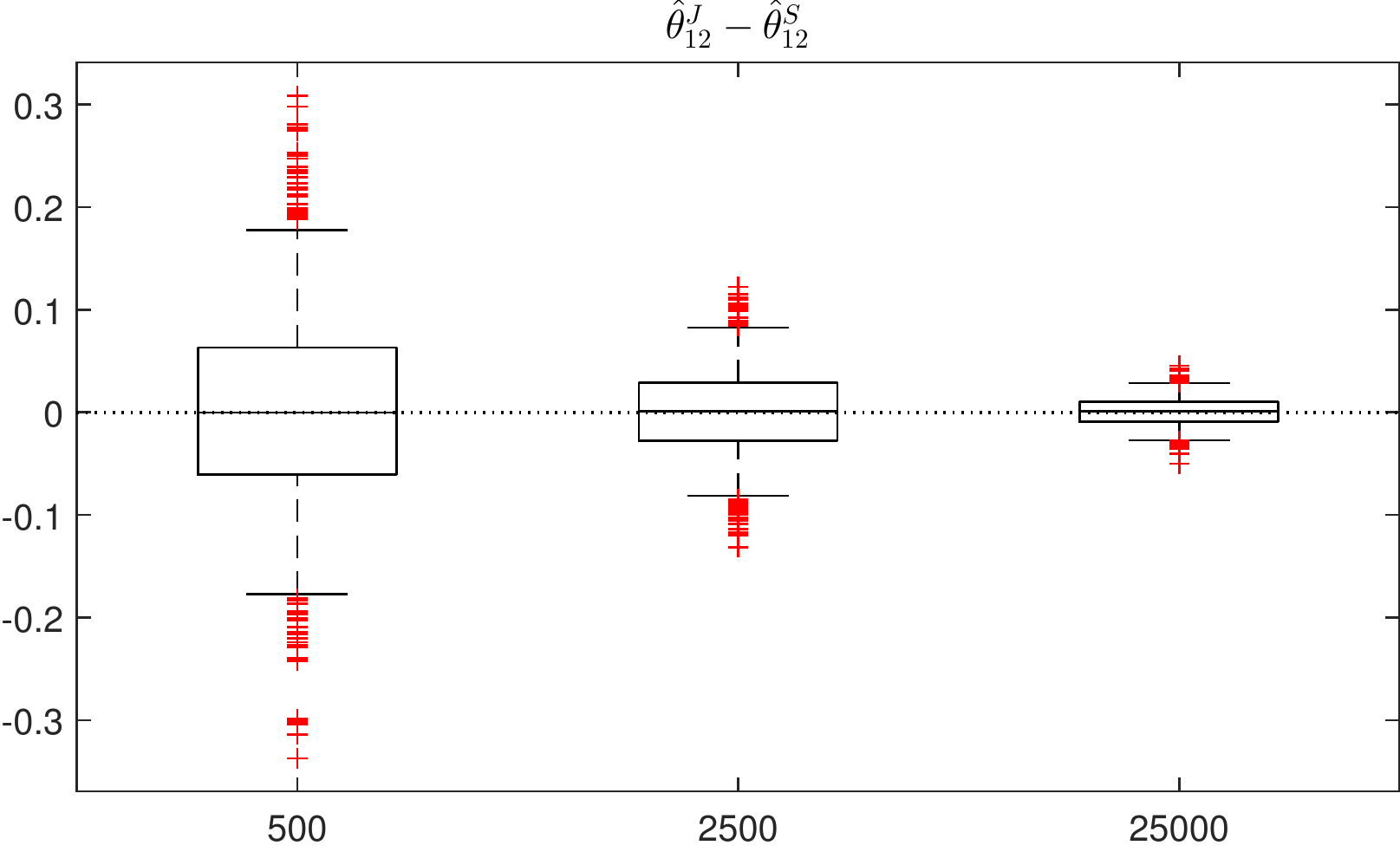}
\end{subfigure}
\smallskip

\begin{subfigure}[t]{0.5\textwidth}
  \centering
    \includegraphics[scale=\scaleit]{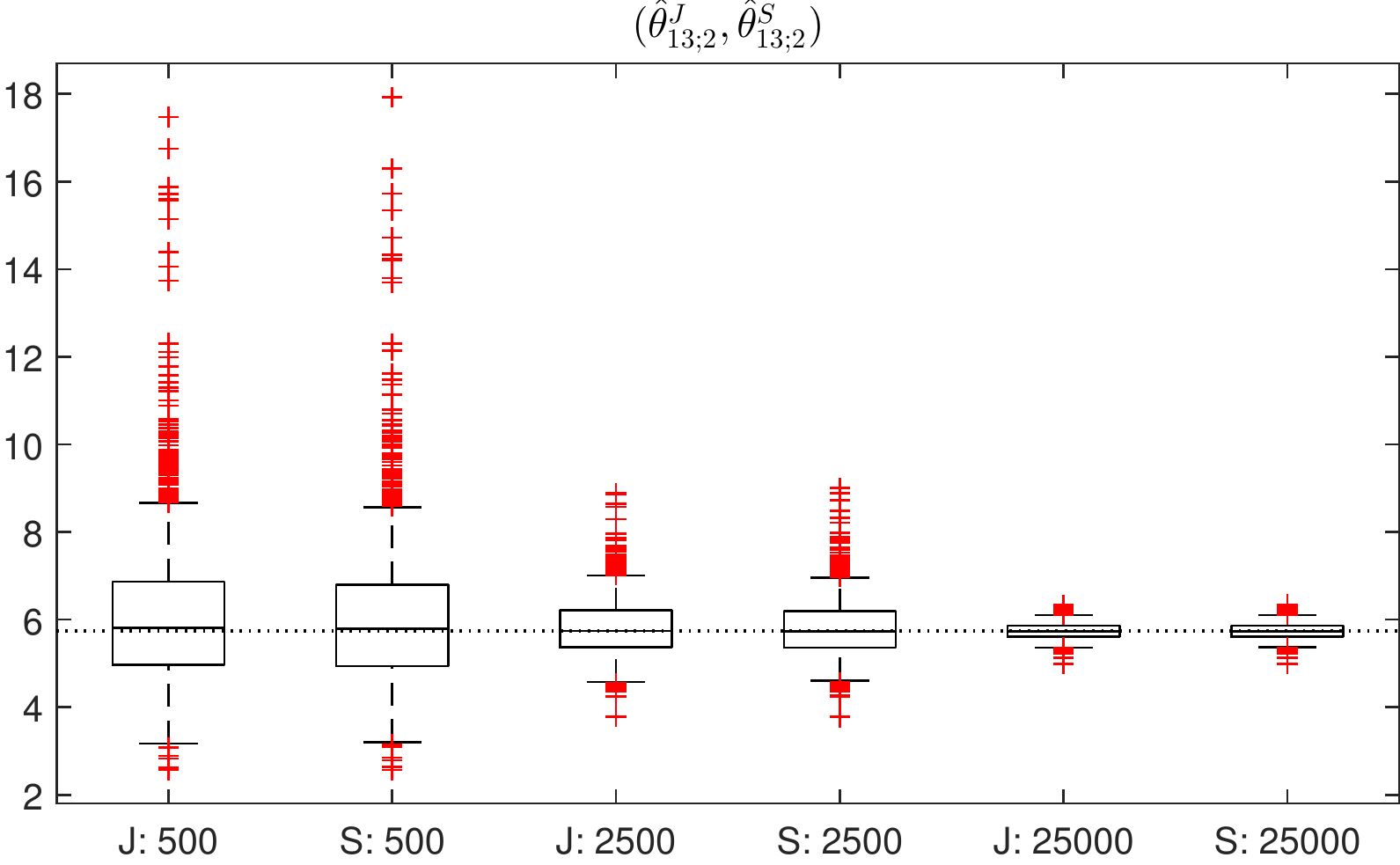}
\end{subfigure}%
\begin{subfigure}[t]{0.5\textwidth}
  \centering
    \includegraphics[scale=\scaleit]{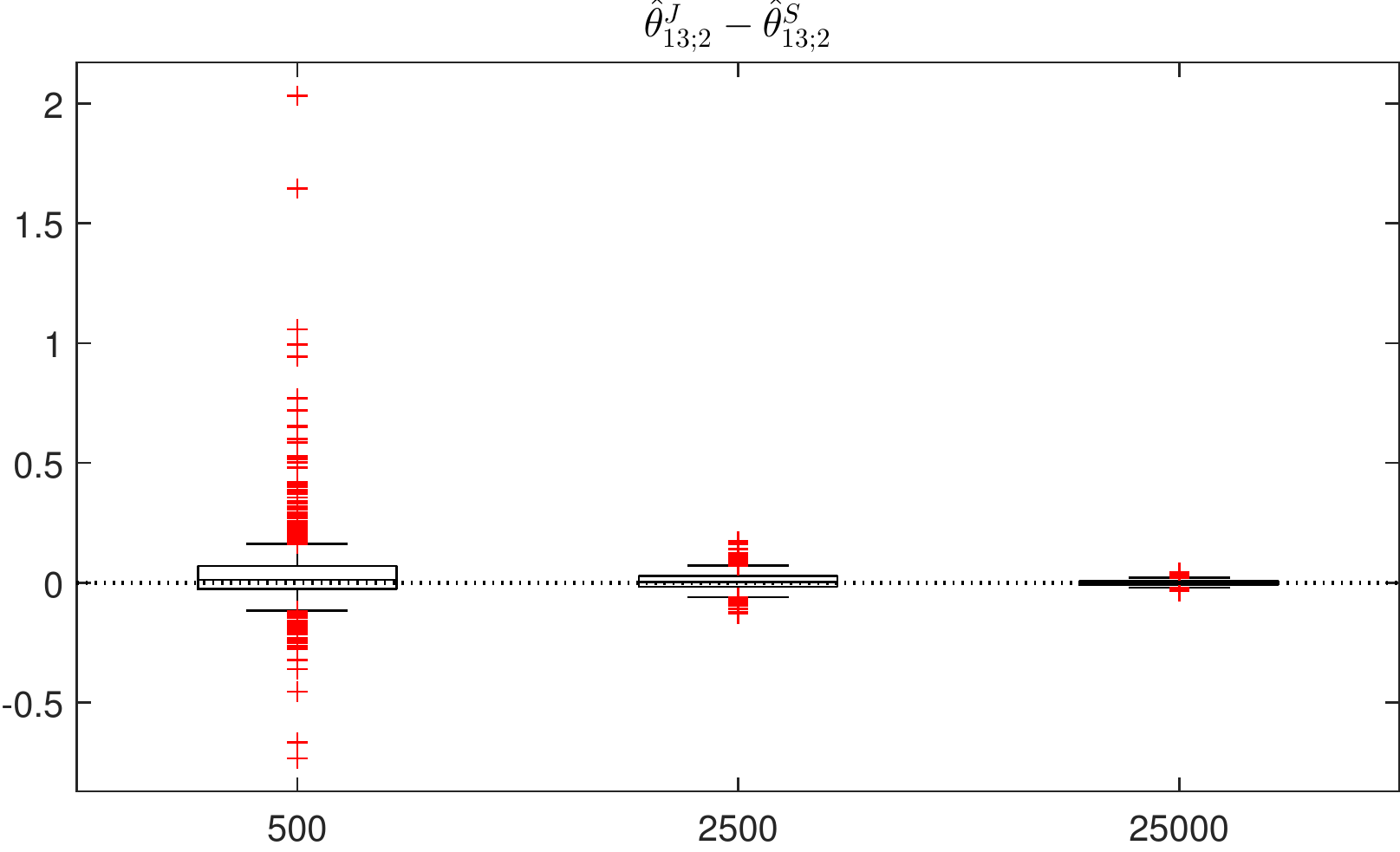}
\end{subfigure}

\captitleno{exA}{pseudo-true parameter}
\label{figexA}
\end{figure}

\autoref{exA} presents a \dgp{} which satisfies the \SA{}, implying 
$\pvc{\theta}=\theta^{\star}$. 
It is the PVC of the three-dimensional Frank copula with Kendall's $\tau$ approximately equal to 0.5. 
\autoref{figexA} shows the corresponding box plots of joint and step-by-step ML estimates and their difference.
The left panel confirms the results of \citet{HobkHaff2012,HobkHaff2013}.
Although the joint ML estimator is more efficient, the loss in efficiency for the step-by-step ML estimator 
is negligible and both estimators converge to the true parameter value. 
Moreover, the right panel of \autoref{figexA} shows that the difference between joint and step-by-step ML estimates is never statistically significant at a 5\% level. 
Since the computational time for a step-by-step ML estimation is much lower than for a joint ML estimation \citep{HobkHaff2012}, the step-by-step ML estimator is very attractive for estimating high-dimensional vine copulas that satisfy the \SA{}.
Moreover, the step-by-step ML estimator is then inherently suited for selecting the pair-copula families in a stepwise manner. 
However, if the \SA{} does not hold for the data generating vine copula, the step-by-step and joint ML estimator can converge to different limits (\autoref{non_consistency}), as the next example demonstrates. 
\begin{myex}[Frank copula]
\label{exB}
Let  $C_{1:3}$ be the Frank copula with dependence parameter $\theta= 5.74$, i.e., $C_{1:3}\neq\pvc{C}_{1:3}$,  and 
$\svcs{C}_{1:3} = 
(C^{\text{Fr}}(\theta_{12}),C^{\text{Fr}}(\theta_{23}),{C}^{\text{P-Fr}}( \theta_{13;2}))$ be the parametric \SVCM{} that is fitted to data generated from $C_{1:3}$.
\end{myex}
\begin{figure}[h!]
\centering

\begin{subfigure}[t]{0.5\textwidth}
  \centering
    \includegraphics[scale=\scaleit]{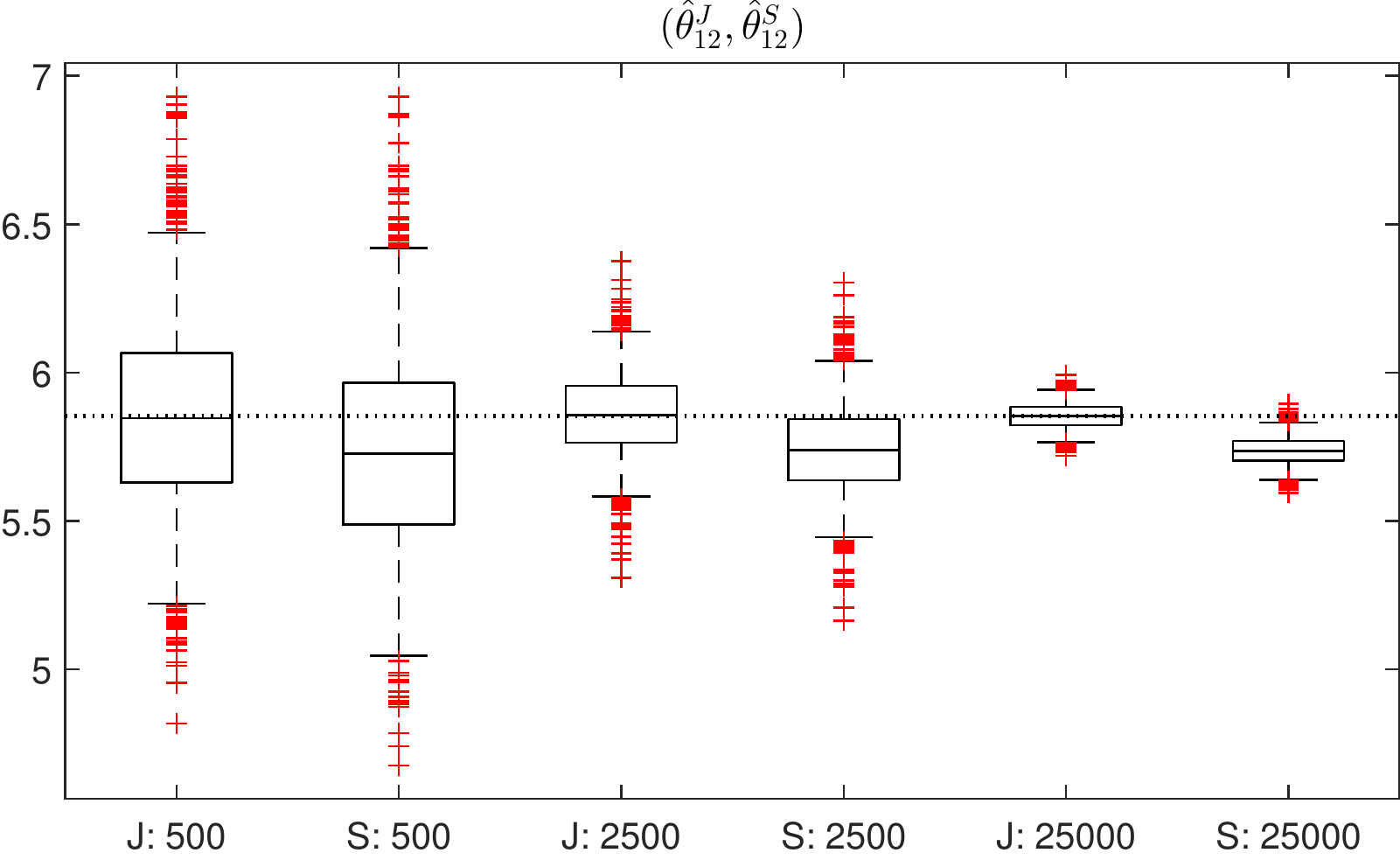}
\end{subfigure}%
\begin{subfigure}[t]{0.5\textwidth}
  \centering
    \includegraphics[scale=\scaleit]{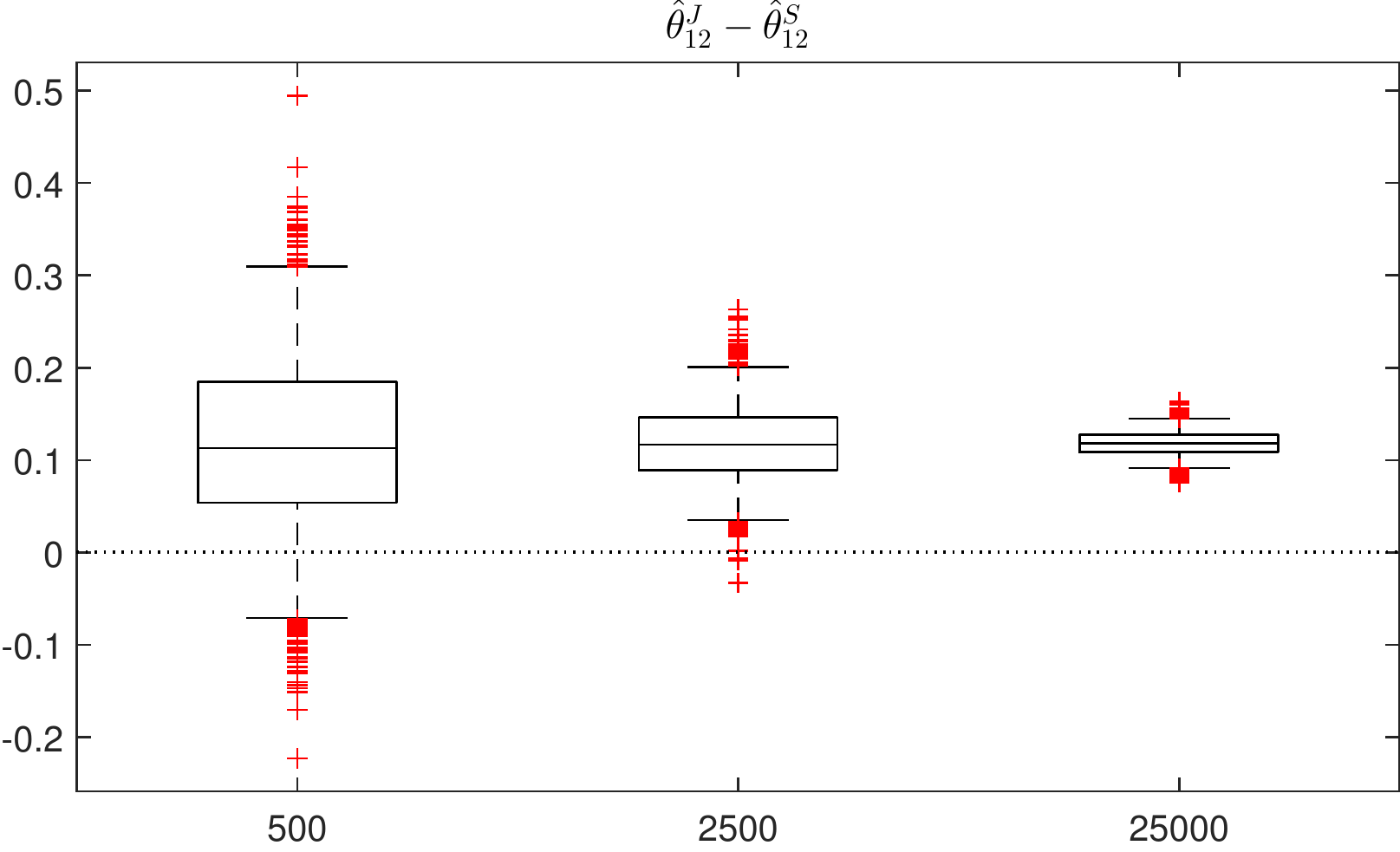}
\end{subfigure}
\smallskip

\begin{subfigure}[t]{0.5\textwidth}
  \centering
    \includegraphics[scale=\scaleit]{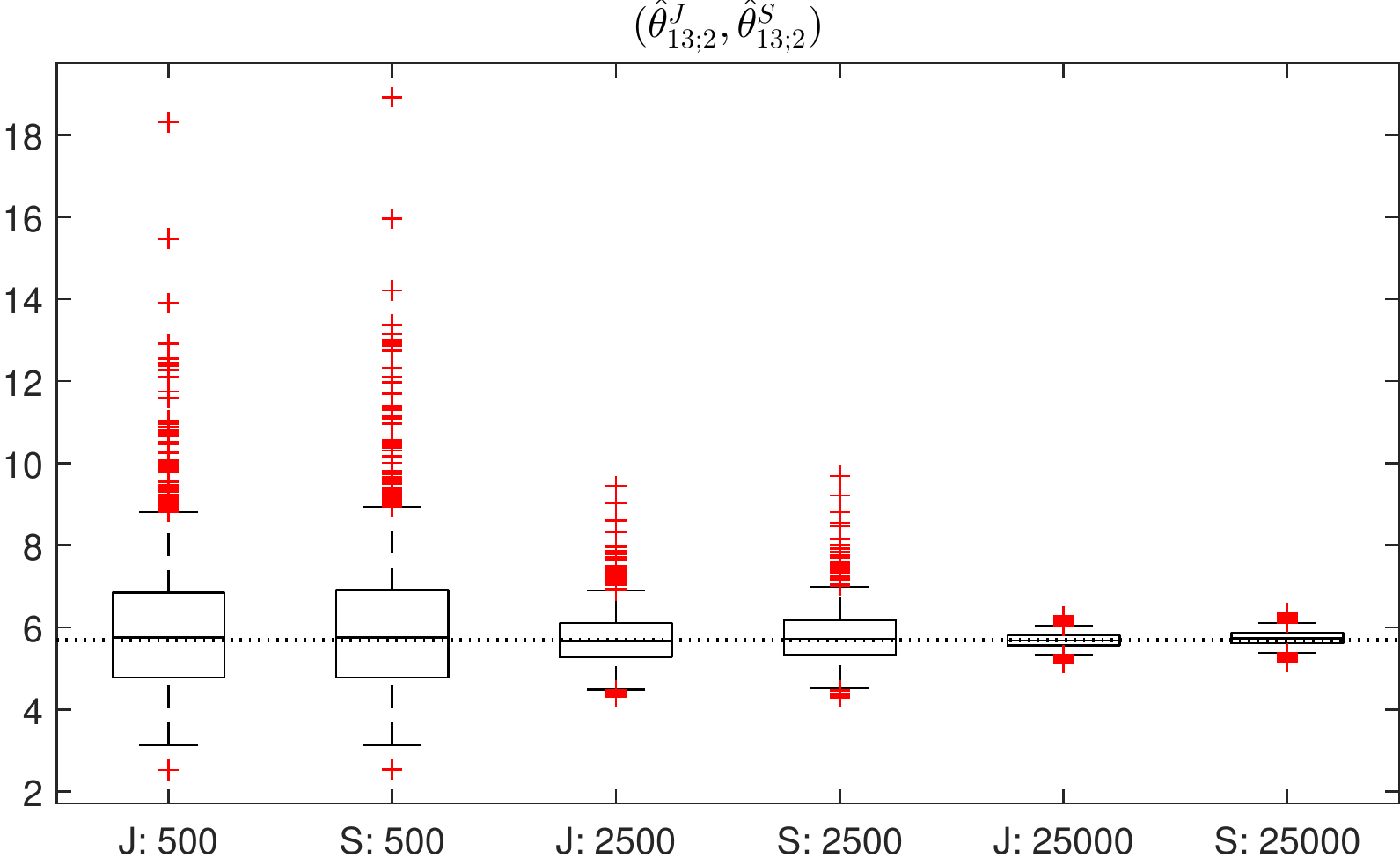}
\end{subfigure}%
\begin{subfigure}[t]{0.5\textwidth}
  \centering
    \includegraphics[scale=\scaleit]{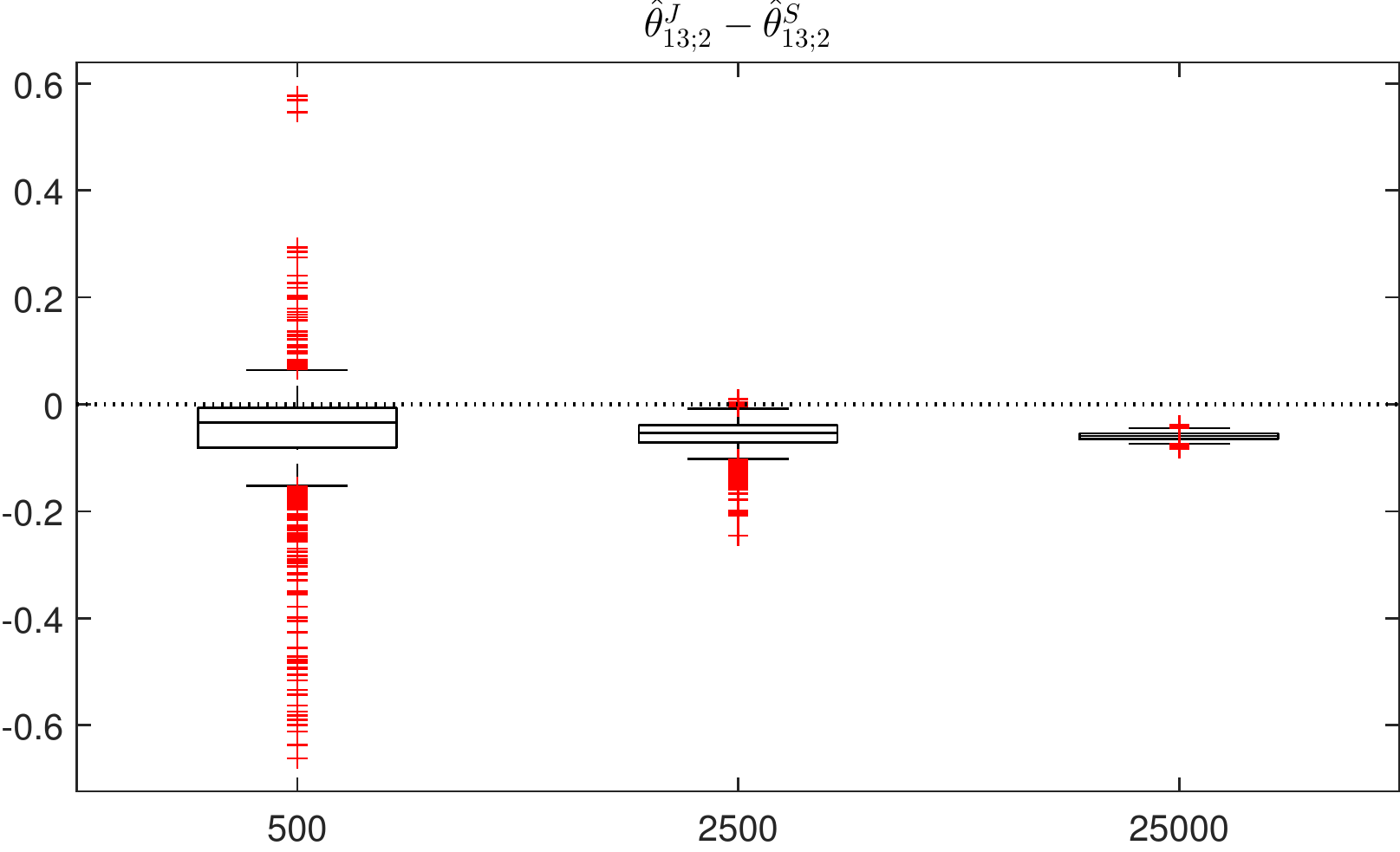}
\end{subfigure}

\captitleno{exB}{pseudo-true parameter}
\label{figexB}

\end{figure}

\autoref{exB} is identical to \autoref{exA}, with the only difference that the conditional copula is varying 
in such a way that the resulting three-dimensional copula is a Frank copula. 
Although the Frank copula does not satisfy the \SA{}, it is pretty close to a copula for which the \SA{} holds, because the variation in the conditional copula is strongly limited for Archimedean copulas (\citet{Mesfioui2008}).
Nevertheless, the right panel of \autoref{figexB}  shows that the step-by-step and joint ML estimates for $\theta_{12}$ are significantly different at the 5\% level if the sample size is 2500 observations. 
The difference between step-by-step and joint ML estimates for $\theta_{13\ps 2}$ is less pronounced, but also highly significant for sample sizes with 2500 observations or more. 
Thus, only in \autoref{exA} the step-by-step ML estimator is a consistent estimator of a simplified vine copula model that 
minimizes the KLD from the underlying copula, whereas the joint ML estimator is a consistent minimizer in both examples. 
A third example where the distance between the data generating copula and the \PVCA{} and thus the difference between the step-by-step and joint ML estimates is more pronounced is given in 
\appref{convergence_appendix}.


\section{Conclusion}
\label{sec_conclusion}
We introduced the partial vine copula (PVC) which is a particular simplified vine copula that coincides with the data generating copula if the \SA{} holds.
The \PVCA{} can be regarded as a generalization of the partial correlation matrix where partial correlations are replaced by $j$-th order partial copulas.
Consequently, it provides a new dependence measure of a $d$-dimensional distribution in terms of $d(d-1)/2$ bivariate unconditional copulas.  
While a higher-order partial copula of the \PVCA{} is related to the partial copula, it does not suffer from the curse of dimensionality and can be estimated for high-dimensional data \citep{Nagler2015}. 
We analyzed to what extent the dependence structure of the underlying distribution is reproduced by the \PVCA{}.  
In particular, we showed that a pair of random variables may be considered as conditionally (in)dependent according to the \PVCA{} although this is not the case for the data generating process.

We also revealed the importance of the \PVCA{} for the modeling of high-dimensional distributions by means of simplified vine copulas (SVCs). Up to now, the estimation of \SVCM{s} has almost always been based on the assumption that the \dgp{} satisfies the \SA{}.
Moreover, the implications that follow if the \SA{} is not true have not been investigated. 
We showed that the \PVCA{} is the \SVCM{} approximation that minimizes the Kullback-Leibler divergence in a stepwise fashion. 
Since almost all estimators of \SVCM{s} proceed sequentially, it follows that, under regularity conditions, many estimators of \SVCM{s} converge to the \PVCA{} also if the \SA{} does not hold. 
However, we also proved that the \PVCA{} may not minimize the Kullback-Leibler divergence from the true copula and thus may  not be the best \SVCM{} approximation in theory. 
Nevertheless, due to the prohibitive computational burden or simply because only a stepwise model specification and estimation is possible, the \PVCA{} is the best feasible \SVCM{} approximation in practice.

The analysis in this paper showed the relative optimality of the \PVCA{} when it comes to approximating multivariate distributions by \SVCM{s}.
Obviously, it is easy to construct (theoretical) examples where the \PVCA{} does not provide a good approximation in absolute terms.
But such examples do not provide any information about the appropriateness of the simplifying assumption in practice.
To investigate whether the \SA{} is true and the \PVCA{} is a good approximation in applications, one can use \autoref{equalppitcpit} to develop tests for the simplifying assumption, see \citet{Kurz2017}.
Moreover, even in cases where the simplifying assumption is strongly violated, an estimator of the \PVCA{} can yield an approximation that is superior to competing approaches.
Recently, it has been demonstrated in \citet{Nagler2015} that the structure of the \PVCA{} can be used to obtain a constrained kernel-density estimator that can be much closer to the data generating process than the classical unconstrained kernel-density estimator, even if the distance between the \PVCA{} and the data generating copula is large.

\section*{Acknowledgements}
We would like to thank Harry Joe and Claudia Czado for comments which helped to improve this paper. 
We also would like to thank Roger Cooke and Ir\`ene Gijbels for interesting discussions on the \SA{}.

\bibliographystyle{model1-num-names}
\bibliography{manuscript}


\section*{Appendix}
\renewcommand{\thesubsection}{A.\arabic{subsection}}
\renewcommand{\thesection}{A}
\subsection{Proof of \autoref{equalppitcpit}}
\label{proof_ppitcpit}
$\pvc{U}_{k|\condset} = U_{k|\condset}\ (\as)
\imp \pvc{U}_{k|\condset}\perp U_{\condset}$ is true because $U_{k|\condset}$ is a CPIT.
For the converse, 
let Let $A:= \times_{k=i+1}^{i+j-1}[0,u_k]$ and
consider
\begin{align}
P(\pvc{U}_{k|\condset}\leq a, U_{\condset}\leq u_{\condset})
&
 = \int_{A}
F_{k|\condset}\big((\pvc{F}_{k|\condset})^
{-1}
(a|t_{\condset})|t_{\condset}\big)
\text{d} C_{{\condset}}(t_{\condset}). \label{refpx2new}
\end{align}
Since $\pvc{U}_{k|\condset}\sim U(0,1)$ it follows that if $\pvc{U}_{k|\condset}\perp U_{\condset}$ then 
$P(\pvc{U}_{k|\condset}\leq a, U_{\condset}\leq u_{\condset})
= aC_{{\condset}}(u_{\condset})$ for all $(a,u_{\condset})\in [0,1]^{j}$.
This implies that
\begin{align*}
P(\pvc{U}_{k|\condset}\leq a, U_{\condset}\leq u_{\condset}) &=
\int_{A}
a\  \!\text{d}C_{{\condset}}(t_{\condset})
\end{align*}
equals the right hand side of \eqref{refpx2new} for all $(a,u_{\condset})\in [0,1]^{j}$. 
It follows that the integrands must be identical (\as) as well and 
$F_{k|\condset}(\pvc{F}_{k|\condset})^
{-1}(a|t_{\condset}) = a $ for all $a\in[0,1]$ and almost every $(u_{\condset})\in [0,1]^{j}$.
Thus $F_{k|\condset}= \pvc{F}_{k|\condset}$ (\as) which is equivalent to
$\pvc{U}_{k|\condset} = U_{k|\condset}\ (\as)$.

\subsection{Proof of \autoref{propmarg}}
\label{prd3}
Let ${C}^{\star}_{1:3}\in \SPCC{C}_3$ be the \SVCM{} given in \autoref{imp_margin}.
We define $C_{1:d}$ as follows.
Let $\condcop_{1,d-1\cs2:d-2} = C^{\star}_{12}, \condcop_{2,d\cs3:d-1} = C^{\star}_{23}$, $\condcop_{1,d\cs 2:d-1}= {D}^{\star}_{1,3\cs 2}$, where $D^{\star}_{1,3\cs 2}$ is the corresponding conditional copula in \autoref{imp_margin}
{and $\condcop_{i,i+j\cs\condset}=E_{k,l}\in {\cal C}_2, (k,l)\in\idxset$ means that
$\condcop_{i,i+j\cs \condset}(a,b|u_{\condset})=E_{k,l}(a,b)$ for all 
$(a,b,u_{\condset})\in[0,1]^{j+1}$.}
Moreover, let
$\condcop_{i,i+j\cs\condset}=C^{\perp}$ for $(i,j)\in \idxset\backslash \{(1,d-2),(2,d-2),(1,d-1)\}$.
The conclusion now follows from \autoref{imp_margin}.

\subsection{Proof of \autoref{cond_indep}}\label{Proof_cond_indep}
{
W.l.o.g. assume that the margins of $X_{1:d}$ are uniform.
Let $C^{FGM_{3}}(u_{1:3};\theta) = \prod_{i=1}^3u_i+\theta\prod_{i=1}^3u_i(1-u_i), |\theta|\leq 1$, be the three-dimensional FGM copula, $d\geq 4$, and $(i,j)\in\idxset$.
It is obvious that $\condcop_{i,i+2\cs i+1} = C^\perp\imp \pvc{C}_{i,i+2\ps i+1}= C^\perp$ is true.
Let $J \in \{2,\ldots,\Mydim-2\}$ be fixed. Assume that $C_{1:\Mydim}$ has 
the following D-vine copula representation of the non-simplified form
\begin{align*}
\condcop_{1,1+J\cs 2:J} &= \partial_{3} C^{FGM_{3}}(u_{1},u_{1+J},u_2;1) \\
\condcop_{2,2+J\cs 3:J+1} &= \partial_{3} C^{FGM_{3}}(u_{2},u_{2+J},u_{1+J};1)
\end{align*}
and $\condcop_{i,i+j\cs \sarg{i}{j}} = C^{\perp}$ for all other $(i,j)\in\idxset$. 
Using the same arguments as in the proof of \autoref{spurious_dep} we obtain
\begin{align*}
\pvc{C}_{i,i+J\ps i+1:J-1} & = C^\perp, 
\quad i=1,2,
\\
\pvc{C}_{1,2+J\ps 2:J+1}  &= 
C^{FGM_2}(1/9).
\end{align*}
This proves that 
$\condcop_{i,i+2\cs i+1} = C^\perp\Leftarrow \pvc{C}_{i,i+2\ps i+1}= C^\perp$ is not true in general and that, for $j\geq 3$, neither the statement 
$\condcop_{i,i+j\cs \condset} = C^\perp \imp \pvc{C}_{i,i+j\ps \condset} = C^\perp$ nor the statement
$\condcop_{i,i+j\cs \condset} = C^\perp \Leftarrow \pvc{C}_{i,i+j\ps \condset} = C^\perp$
 is true in general.
}

\subsection{Proof of \autoref{spurious_dep}}
\label{Derive_spurious_dep}

We show a more general result and set 
$\condcop_{i,i+2\cs i+1}(u_i,u_{i+2}|u_{i+1})  = C^{FGM_{2}}(u_i,u_{i+2};g(u_{i+1}))$ in \eqref{uncondfgm}
where $g\colon[0,1]\to [-1,1]$ is a non-constant measurable function such that 
\begin{align}
\label{gsym}
\forall u\in [0.5,1]\colon g(0.5+u)=-g(0.5-u).
\end{align}

For $i=1,2,3,$ the copula in the second tree of the PVC is given by
\begin{align*}
\pvc{C}_{i,i+2\ps i+1}(a,b) &= \mathbb{P}(U_{i|i+1} \leq a, U_{i+2|i+1} \leq b)
= \int_{[0,1]} \condcop_{i,i+2\cs i+1}(a,b|u_{i+1}) \text{d}u_{i+1} \\
& \ \ \RefEqual{\eqref{uncondfgm}}\ \  ab\big(1 + (1-a)(1-b) \int_{[0,1]} g(u_{i+1}) \text{d}u_{i+1}\big) 
\ \ \RefEqual{\eqref{gsym}}\ \ ab,\tagging\label{eqPartial}
\end{align*}
which is the independence copula.
For $i=1,2, k=i,i+3$, the true CPIT of $U_k$ w.r.t. $U_{i+1:i+2}$ is a function of $U_{i+1:i+2}$ because 
\begin{align*}
U_{i|i+1:i+2} &= U_{i}[1+g(U_{i+1})(1-U_{i})(1-2U_{i+2})]\tagging\label{eqCPIT1_1},\\
U_{i+3|i+1:i+2} &=  U_{i+3}[1+g(U_{i+2})(1-U_{i+3})(1-2U_{i+1})].\tagging\label{eqCPIT1_2}
\end{align*}
However, for $i=1,2, k=i,i+3$, the PPIT of $U_k$ w.r.t. $U_{i+1:i+2}$ is not a function of $U_{i+1:i+2}$ because
\begin{align*}
\pvc{U}_{i|i+1:i+2} &= \pvc{F}_{i|i+1:i+2}(U_{i}|U_{i+1:i+2}) 
= F_{U_{i|i+1}|U_{i+2|i+1}}(U_{i|i+1}|U_{i+2|i+1}) \\
&= \partial_2 \pvc{C_{i,i+2\ps i+1}}(U_{i|i+1},U_{i+2|i+1}) 
\ \  \RefEqual{\eqref{eqPartial}}\ \ U_{i|i+1} \ \
\RefEqual{\eqref{eqCPIT}} \ \ U_{i} \tagging\label{eqPCPIT1_1},
\shortintertext{and, by symmetry, }
\pvc{U}_{i+3|i+1:i+2}& = U_{i+3}. \tagging\label{eqPCPIT1_2}
\end{align*}
For $i=1,2$, the joint distribution of these {first-order} PPITs is a copula in the third tree of the PVC which is given by
\begin{align}
\pvc{C}_{i,i+3\ps i+1:i+2}(a,b) &= 
\mathbb{P}(\pvc{U}_{i|i+1:i+2} \leq a, \pvc{U}_{i+3|i+1:i+2} \leq b)
\quad\ \ \RefEqual{\eqref{eqPCPIT1_1},\eqref{eqPCPIT1_2}}\quad \ \ \mathbb{P}(U_{i} \leq a, U_{i+3} \leq b) 
= C_{i,i+3}(a,b) \tagging\label{uncondmargfgm}\\
& \RefEqual{\eqref{uncondfgm2}}\quad \int_{[0,1]^2} F_{i|i+1:i+2}(a|u_{i+1:i+2})F_{i+3|i+1:i+2}(b|u_{i+1:i+2}) \text{d}u_{i+1:i+2}\notag \\
&\RefEqual{\eqref{eqCPIT1_1},\eqref{eqCPIT1_2}}\quad \ \
ab[1+ (1-a)(1-b)\int_{[0,1]}g(u_{i+1})(1-2u_{i+1})\text{d}u_{i+1} \int_{[0,1]}g(u_{i+2})(1-2u_{i+2})
\text{d}u_{i+2} ]\notag \\
&= ab[1+ \theta (1-a)(1-b)] = C^{FGM_{2}}(\theta),\notag
\end{align}
where $\theta := 4 (\int_{[0,1]} ug(u) \text{d}u)^2>0$, by the properties of $g$.
Thus, a copula in the third tree of the PVC is a bivariate FGM copula whereas the true conditional copula is the independence copula. 

The CPITs of $U_1$ or $U_5$ w.r.t. $U_{2:4}$ are given by 
\begin{align*}
U_{1|2:4} &= F_{1|2:4}(U_{1}|U_{2:4}) 
= \partial_2 \condcop_{14\cs 2:3}(U_{1|2:3},U_{4|2:3}|U_{2:3}) 
\ \ \RefEqual{\eqref{uncondfgm2}} \ \
 U_{1|2:3}  \\
&\RefEqual{\eqref{eqCPIT1_1}}\ \ U_{1}[1+g(U_{2})(1-U_{1})(1-2U_{3})],
\tagging\label{eqCPIT2_1}
\\
U_{5|2:4} &=\ \   U_{5}[1+g(U_{4})(1-U_{5})(1-2U_{3})],\tagging\label{eqCPIT2_2}
\end{align*}
whereas the corresponding {second-order PPITs} are given by
\begin{align*}
\pvc{U}_{1|2:4} &= \pvc{F}_{1|2:4}(U_{1}|U_{2:4}) 
= F_{\pvc{U_{1|2:3}}|\pvc{U_{4|2:3}}}(\pvc{U_{1|2:3}}|\pvc{U_{4|2:3}}) \quad\ \ \RefEqual{\eqref{eqPCPIT1_1},\eqref{eqPCPIT1_2}}\quad\ \ F_{U_{1}|U_{4}}(U_{1}|U_{4}) \\
&= U_{1|4} \tagging\label{eqPCPIT2_1} 
= \partial_2 C_{14}(U_{1},U_{4}) 
\quad\RefEqual{\eqref{uncondmargfgm}}\quad U_{1}[1+\theta(1-U_{1})(1-2U_{4})],
\\
\pvc{U_{5|2:4}} &=  U_{5}[1+\theta(1-U_{5})(1-2U_{2})]. \tagging\label{eqPCPIT2_2}
\end{align*}

For the copula in the fourth tree of the PVC it holds
\allowdisplaybreaks
\begin{align*}
\pvc{C}_{15\ps 2:4}(a,b) &= \mathbb{P}(\pvc{U}_{1|2:4} \leq a, \pvc{U}_{5|2:4} \leq b) 
\quad \;\RefEqual{\eqref{eqPCPIT2_1},\eqref{eqPCPIT2_2}}\quad \; \mathbb{P}(U_{1|4} \leq a, U_{5|2} \leq b)
= \mathbb{P}(U_{1} \leq F_{1|4}^{-1}(a|U_{4}), U_{5} \leq F_{5|2}^{-1}(b|U_{2})) \\
&= \int_{[0,1]^3} F_{15|2:4}(F_{1|4}^{-1}(a|u_{4}),F_{5|2}^{-1}(b|u_{2})|u_{2:4})c_{2:4}(u_{2:4}) \text{d}u_{2:4} \\
&= \int_{[0,1]^3} \condcop_{15\cs 2:4}(F_{1|2:4}(F_{1|4}^{-1}(a|u_{4})|u_{2:4}),F_{5|2:4}(F_{5|2}^{-1}(b|u_{2})|u_{2:4})|u_{2:4})c_{2:4}(u_{2:4}) \text{d}u_{2:4} \\
& \RefEqual{\eqref{uncondfgm3}} \int_{[0,1]^3} F_{1|2:4}(F_{1|4}^{-1}(a|u_{4})|u_{2:4})F_{5|2:4}(F_{5|2}^{-1}(b|u_{2})|u_{2:4})c_{2:4}(u_{2:4}) \text{d}u_{2:4} \\
&\RefEqual{\eqref{eqCPIT2_1},\eqref{eqCPIT2_2}}\quad \int_{[0,1]^3} F_{1|2:3}(F_{1|4}^{-1}(a|u_{4})|u_{2:3})F_{5|3:4}(F_{5|2}^{-1}(b|u_{2})|u_{3:4})c_{2:4}(u_{2:4}) \text{d}u_{2:4} \\
&\RefEqual{\eqref{eqCPIT1_1},\eqref{eqCPIT1_2}}\quad \int_{[0,1]^3} F_{1|4}^{-1}(a|u_{4})[1+g(u_{2})(1-F_{1|4}^{-1}(a|u_{4}))(1-2u_{3})] \\ &\qquad\times F_{5|2}^{-1}(b|u_{2})[1+g(u_{4})(1-F_{5|2}^{-1}(b|u_{2}))(1-2u_{3})] \times [1+g(u_3)(1-2u_{2})(1-2u_{4})] \text{d}u_{2:4}\\
&= \int_{[0,1]^2} F_{1|4}^{-1}(a|u_{4})F_{5|2}^{-1}(b|u_{2}) \left[1+\int_{[0,1]} (1-2u_{3})^2 \text{d}u_{3} (1-F_{1|4}^{-1}(a|u_{4}))(1-F_{5|2}^{-1}(b|u_{2})) g(u_{4}) g(u_{2}) \right. \\ &\qquad+ \int_{[0,1]} (1-2u_{3})g(u_{3}) \text{d}u_{3} (1-F_{5|2}^{-1}(b|u_{2}))(1-2u_{2})(1-2u_{4}) g(u_{4}) \\ &\qquad+ \left. \int_{[0,1]} (1-2u_{3})g(u_{3}) \text{d}u_{3} (1-F_{1|4}^{-1}(a|u_{4}))(1-2u_{4})(1-2u_{2})g(u_{2})\right] \text{d}u_{2}\text{d}u_{4},
\intertext{where we used that $\int_{[0,1]} (1-2u_{3}) \text{d}u_{3} = 0$, $\int_{[0,1]} g(u_{3}) \text{d}u_{3} \RefEqual{\eqref{gsym}} 0$ and $\int_{[0,1]} (1-2u_{3})^2 g(u_{3}) \text{d}u_{3} \RefEqual{\eqref{gsym}}0$. 
By setting $\gamma := -2 \int_{[0,1]} ug(u) \text{d}u$ we can write the copula function as}
\pvc{C}_{15\ps 2:4}(a,b) &= \int_{[0,1]^2} F_{1|4}^{-1}(a|u_{4})F_{5|2}^{-1}(b|u_{2}) \left[1+\frac{1}{3}(1-F_{1|4}^{-1}(a|u_{4}))(1-F_{5|2}^{-1}(b|u_{2})) g(u_{4}) g(u_{2}) \right. \\ &\qquad+ \gamma (1-F_{5|2}^{-1}(b|u_{2}))(1-2u_{2})(1-2u_{4}) g(u_{4}) \\ &\qquad+ \left. \gamma (1-F_{1|4}^{-1}(a|u_{4}))(1-2u_{4})(1-2u_{2})g(u_{2})\right] \text{d}u_{2}\text{d}u_{4}\\
&= \left(\int_{0}^{1} F_{1|4}^{-1}(a|u_{4})\text{d}u_{4}\right) \left(\int_{0}^{1}F_{5|2}^{-1}(b|u_{2}) \text{d}u_{2}\right) \\
&\qquad+ \gamma \left(\int_{0}^{1}(1-2u_{2})(F_{5|2}^{-1}(b|u_{2})-(F_{5|2}^{-1}(b|u_{2}))^2)\text{d}u_{2}\right) \left(\int_{0}^{1} (1-2u_{4}) g(u_{4}) F_{1|4}^{-1}(a|u_{4}) \text{d}u_{4}\right) \\
&\qquad+ \gamma \left(\int_{0}^{1}(1-2u_{4})(F_{1|4}^{-1}(a|u_{4})-(F_{1|4}^{-1}(a|u_{4}))^2)\text{d}u_{4}\right) \left(\int_{0}^{1} (1-2u_{2}) g(u_{2}) F_{5|2}^{-1}(b|u_{2}) \text{d}u_{2}\right) \\
&\qquad+ \frac{1}{3}\left(\int_{0}^{1}g(u_{4})(F_{1|4}^{-1}(a|u_{4})-(F_{1|4}^{-1}(a|u_{4}))^2)\text{d}u_{4}\right) \left(\int_{0}^{1}g(u_{2})(F_{5|2}^{-1}(b|u_{2})-(F_{5|2}^{-1}(b|u_{2}))^2) \text{d}u_{2}\right)
\end{align*}
If $(U,V) \sim C^{FGM_{2}}(\theta)$,  the quantile function is given by (cf. \citet{Remillard2013})
\begin{align*}
F^{-1}_{U|V}(u|v) 
&= \frac{1+h(v)-\sqrt{(1+h(v))^2-4h(v)u}}{2h(v)},
\intertext{with $h(v):= \theta (1-2v)$, which implies}
\frac{\partial}{\partial u} F_{U|V}^{-1}(u|v) &= \frac{1}{\sqrt{(1+h(v))^2-4h(v)u}} =:G(u,v), \tagging\label{eqFGMderiv}
\intertext{and}
\frac{\partial}{\partial u} (F^{-1}_{U|V}(u|v))^2 &= \frac{1}{h(v)}\left[(1+h(v))G(u,v)-1\right]. \tagging\label{eqFGM2deriv}
\end{align*}
For the density of the copula in the fourth tree of the PVC it follows
\begin{align*}
\pvc{c}_{15\ps 2:4}(a,b) &= \frac{\partial^2}{\partial a \partial b} \pvc{C}_{15\ps 2:4}(a,b)\\
&\RefEqual{\eqref{eqFGMderiv},\eqref{eqFGM2deriv}}\quad\; \left(\int_{0}^{1} G(a,u_{4})\text{d}u_{4}\right) \left(\int_{0}^{1} G(b,u_{2}) \text{d}u_{2}\right) \\
&\qquad+ \frac{1}{\gamma} \left(1- \int_{0}^{1} G(b,u_{2}) \text{d}u_{2}\right) \left(\int_{0}^{1}  (1-2u_{4}) g(u_{4})G(a,u_{4}) \text{d}u_{4}\right) \\
&\qquad+ \frac{1}{\gamma} \left(1- \int_{0}^{1} G(a,u_{4}) \text{d}u_{4}\right) \left(\int_{0}^{1}  (1-2u_{2}) g(u_{2})G(b,u_{2}) \text{d}u_{2}\right) \\
&\qquad+ \frac{1}{3}\left(\int_{0}^{1} \frac{g(u_{4})}{h(u_{4})}\left[1-G(a,u_{4})\right] \text{d}u_{4}\right) \left(\int_{0}^{1} \frac{g(u_{2})}{h(u_{2})}\left[1-G(b,u_{2})\right] \text{d}u_{2}\right) \\
&= \frac{1}{4\theta^2} \log(\sigma(a)) \log(\sigma(b)) + \frac{1}{\gamma} \left(1- \frac{1}{2\theta}\log(\sigma(b))\right) \left(\int_{0}^{1}  (1-2u_{4}) g(u_{4})G(a,u_{4}) \text{d}u_{4}\right) \\
&\qquad+ \frac{1}{\gamma} \left(1- \frac{1}{2\theta}\log(\sigma(a))\right) \left(\int_{0}^{1}  (1-2u_{2}) g(u_{2})G(b,u_{2}) \text{d}u_{2}\right) \\
&\qquad+ \frac{1}{3}\left(\int_{0}^{1} \frac{g(u_{4})}{h(u_{4})}\left[1-G(a,u_{4})\right] \text{d}u_{4}\right) \left(\int_{0}^{1} \frac{g(u_{2})}{h(u_{2})}\left[1-G(b,u_{2})\right] \text{d}u_{2}\right),
\shortintertext{where}
\sigma(i) &= 
\left(\frac{\sqrt{(1+\theta)^2-4\theta i}+1-2i +\theta}{\sqrt{(1-\theta)^2+4\theta i}+1-2i -\theta}\right) \quad \text{ for $i\in \{a,b\}$}.
\end{align*}
If we set $g(u) := 1-2u$, then $\theta = 1/9$ and $\gamma = 1/3$, and we get
\begin{align*}
&\pvc{c}_{15\ps 2:4}(a,b)
= \frac{81}{4} \prod_{i=a,b}\log (s(i))+ 27\prod_{i=a,b}
 \left(1- \frac{81}{4}\log (s(i))\right) 
\\ & \quad+ \frac{2187}{2}\!\!
 \sum_{(i,j)\in I_{a,b}}\!\!
\left(1- \frac{9}{2}\log (s(i))\right)
\Bigg[ (6j^2-6j+1) \log (s(j)) + \frac{1}{9}\left((6j-\frac{26}{9})\sqrt{25-9j} - (6j-\frac{28}{9})\sqrt{16+9j}\right) \Bigg] \\ 
\shortintertext{where}
&s(i) = \frac{\sqrt{25-9i}+5-9i}{\sqrt{16+9i}+4-9i} \quad \text{ for $i\in \{a,b\}$ and $I_{a,b} := \lbrace(a,b),(b,a)\rbrace$}.
\end{align*}
Evaluating the density shows that 
$\pvc{C}_{15\ps 2:4}$ is not the independence copula.

\newcommand{\condsetn}{\sarg{i}{n}} 
\newcommand{\Gfunn}[2]{G_{#1|\condsetn}(#2|u_{\condsetn})}
\subsection{Proof of \autoref{hop}}
\label{proof_seqkl}
The KLD related to tree $j$, $D_{KL}^{(j)}({\tree{j}} ({{\cal T}}_{1:j-1}))$, is minimized when the negative cross entropy related to tree $j$ is maximized.
The negative cross entropy related to tree $j$ is given by
\begin{align*}
H^{(j)}({\tree{j}} ({{\cal T}}_{1:j-1})) 
& := \sum_{i=1}^{d-j} \mathbb{E}
\left[\log 
{
\svcs{c}_{i,i+j\ps \condset}(\svcs{F}_{i|\condset}(U_i|U_{\condset}),
\svcs{F}_{i+j|\condset}(U_{i+j}|U_{\condset}))
}
\right]
\\
& = : \sum_{i=1}^{d-j} H_i^{(j)}(\svcs{c}_{i,i+j\ps \condset},\svcs{F}_{i|\condset},
\svcs{F}_{i+j|\condset}).
\end{align*}
Obviously, to maximize $H^{(j)}({\tree{j}} ({{\cal T}}_{1:j-1}))$ w.r.t. ${\tree{j}}$
we can maximize each 
$H_i^{(j)}(\svcs{c}_{i,i+j\ps \condset},\svcs{F}_{i|\condset},
\svcs{F}_{i+j|\condset})$ individually for all $i=1,\ldots,d-j$.
If $j=1$, then 
\begin{align*}
H_i^{(j)}(\svcs{c}_{i,i+j\ps \condset},\pvc{F}_{i|\condset},
\pvc{F}_{i+j|\condset}) & = 
\mathbb{E}
\left[
\log \frac{c_{i,i+1}(U_i,U_{i+1})}{\svcs{c}_{i,i+1}(U_i,U_{i+1})}
\right]
\end{align*}
which is maximized for 
$\svcs{C}_{i,i+1}= C_{i,i+1}$ by Gibbs' inequality.
Thus, if $j=1$, then 
\begin{align}
\label{toproofseq}
\std{\arg\min}{\tree{j}\in\allvines{j}}\quad\!
D_{KL}^{(j)}(\tree{j} (\mpartreeseq{j-1})) 
&= \pvc{{\cal T}}_{j}.
\end{align}
To show that \eqref{toproofseq} holds for $j\geq 2$ we use induction. 
Assume that 
\begin{align*}
\std{\arg\min}{\tree{j}\in\allvines{j}}\quad\!
D_{KL}^{(j)}(\tree{j} (\mpartreeseq{j-1})) 
&= \pvc{{\cal T}}_{j}.
\end{align*}
holds for $1\leq j \leq d-2$.
To minimize the KLD related to tree $j+1 =:n$ w.r.t. $\tree{n}$, conditional on 
$ {{\cal T}}_{1:n-1} =\mpartreeseq{n-1}$,
 we have to maximize the negative cross entropy which is maximized if 
\begin{align*}
&H_i^{(n)}(\svcs{c}_{i,i+n\ps \condsetn},\pvc{F}_{i|\condsetn},
\pvc{F}_{i+n|\condsetn})
\\
& = \mathbb{E}\left[\log\svcs{c}_{i,i+n\ps \condsetn}
\big(\pvc{F}_{i|\condsetn}(U_i|U_{\condsetn}),
\pvc{F}_{i+n|\condsetn}(U_{i+n}|U_{\condsetn})\big)
\right]
\end{align*}
is maximized for all $i=1,\ldots,d-n$.
Using the substitution 
$u_i = (\pvc{F}_{i|\condsetn})^{-1}(t_{i}|u_{\condsetn}) = \Gfunn{i}{t_{i}}$
and
$u_{i+n} = (\pvc{F}_{i+n|\condsetn})^{-1}(t_{i+n}|u_{\condsetn})
=\Gfunn{i+n}{t_{i+n}}$,
we obtain
{
\begin{align*}
&H_i^{(n)}(\svcs{c}_{i,i+n\ps \condsetn},\pvc{F}_{i|\condsetn},
\pvc{F}_{i+n|\condsetn}) =
\int_{[0,1]^{n+1}}\log \svcs{c}_{i,i+n\ps \condsetn}(t_{i},t_{i+n}) 
\\
&\quad \times
\condden_{i,i+n\cs\condsetn}
\Big(
F_{i|\condsetn}\big(\Gfunn{i}{t_{i}}
\big|u_{\condsetn}\big),
F_{i|\condsetn}\big(\Gfunn{i+n}{t_{i+n}}
\big|u_{\condsetn}\big)
\Big|u_{\condsetn} \Big)
\\
& \quad \times
\frac{\prod_{k=i,i+n}f_{k|\condsetn}\big(\Gfunn{k}{t_k}\big|u_{\condsetn}\big)}
{\prod_{k=i,i+n}\pvc{f}_{k|\condsetn}\big(\Gfunn{k}{t_k}\big|u_{\condsetn}\big)}
c_{\condsetn}(u_{\condsetn})\text{d}u_{\condsetn} \text{d}t_{i} \text{d}t_{i+n}
\\
&=\int_{[0,1]^{2}}\log \svcs{c}_{i,i+n\ps \condsetn}(t_{i},t_{i+n}) 
\\
& \quad\times
\Big(
\int_{[0,1]^{n-1}}
\condden_{i,i+n\cs\condsetn}
\Big(
F_{i|\condsetn}\big(\Gfunn{i}{t_{i}}
\big|u_{\condsetn}\big),
F_{i|\condsetn}\big(\Gfunn{i+n}{t_{i+n}}
\big|u_{\condsetn}\big)
\Big|u_{\condsetn} \Big)
\\
& \quad \times
\frac{\prod_{k=i,i+n}f_{k|\condsetn}\big(\Gfunn{k}{t_k}\big|u_{\condsetn}\big)}
{\prod_{k=i,i+n}\pvc{f}_{k|\condsetn}\big(\Gfunn{k}{t_k}\big|u_{\condsetn}\big)}
c_{\condsetn}(u_{\condsetn})  \text{d}u_{\condsetn}
\Big)\text{d}t_{i} \text{d}t_{i+n}
\\
&=\int_{[0,1]^{2}}\log \svcs{c}_{i,i+n\ps \condsetn}(t_{i},t_{i+n}) 
\pvc{c}_{i,i+n\ps \condsetn}(t_{i},t_{i+n})\text{d}t_{i} \text{d}t_{i+n},
\end{align*}
}
which is maximized for $\svcs{c}_{i,i+n\ps \condsetn}=
\pvc{c}_{i,i+n\ps \condsetn}=\pvc{c}_{i,i+(j+1)\ps \sarg{i}{j+1}}$ by Gibbs' inequality.

\subsection{Proof of \autoref{prop_kl}}
\label{proof_prop_kl}
Equation \eqref{gkl1} is obvious, since 
$\pvc{C}_{1:d} $ is the data generating process. 
Equation \eqref{gkl3} immediately follows from the equations \eqref{sbs3} and \eqref{gkl2}.
Using {the same arguments as in \appref{prd3}}, the validity of \eqref{gkl2} for $d=3$ implies the validity of \eqref{gkl2} for $d\geq 3$.
{
However, even for $d=3$, the KLD is a triple integral and does not exhibit an analytical expression if the data generating process is a non-simplified vine copula. 
Thus, the hard part is to show that there exists a data generating copula which does not satisfy the simplifying assumption and for which the \PVCA{} does not minimize the KLD.
}
We prove equation \eqref{gkl2} for $d=3$ by means of the following example.
\renewcommand{\tpar}{\pvc{\theta}_{13; 2}}
\begin{myex}
\label{KLex}
Let $g\colon[0,1]\to[-1,1]$ be a measurable function.
Consider the data generating process
\begin{align*}
C_{\idxlist}(u_{\idxlist}) & = \int_0^{u_\idxo}
C^{FGM_{2}}\big(u_\idxz,u_\idxt;g(z)\big)\text{d}z,
\end{align*}
i.e., the two unconditional bivariate margins $(C_{12},C_{23})$ are independence copulas and the conditional copula is a FGM copula  with varying parameter
$g(u_\idxo)$. 
The first-order partial copula is also a FGM copula given by
\begin{align*}
\svcs{C}_{\idxz\idxt\ps \idxo}(u_\idxz,u_\idxt;\tpar) & = u_\idxz u_\idxt
[1+\tpar (1-u_\idxz)(1-u_\idxt)], \quad \tpar :=\int_0^1g(u_\idxo)\text{d}u_\idxo.
\end{align*}
We set
$\svcs{C}_{\idxo\idxt} = C_{\idxo\idxt}, \svcs{C}_{\idxz\idxt\ps \idxo} = \pvc{C}_{\idxz\idxt\ps \idxo}$, and specify 
a parametric copula 
$\svcs{C}_{\idxz\idxo}(\tmarg), \tmarg\in\Theta_{12}\subset \R,$ 
with conditional cdf $\svcs{F}_{\idxz|\idxo}(u_\idxz|u_\idxo;\tmarg)$ and such that $\svcs{C}_{\idxz\idxo}(0)$ corresponds to the independence copula. 
Thus, $(\svcs{C}_{\idxz\idxo}(0),C_{\idxo\idxt},\pvc{C}_{\idxz\idxt\ps \idxo}) = 
(C_{\idxz\idxo},C_{\idxo\idxt},\pvc{C}_{\idxz\idxt\ps \idxo})$.
We also assume that $\svcs{c}_{\idxz\idxo}(u_1,u_2;\tmarg)$ and $\partial_{\tmarg}\svcs{c}_{\idxz\idxo}(u_1,u_2;\tmarg)$ are both continuous on $(u_1,u_2,\tmarg)\in(0,1)^2\times\Theta_{12}$.
\end{myex}

We now derive necessary and sufficient conditions such that 
\begin{align*}
\KL{C_{\idxlist}}{\svcs{C}_{\idxz\idxo}(\tmarg),C_{\idxo\idxt},\pvc{C}_{\idxz\idxt\ps \idxo}}:=\KL{C_{\idxlist}}
{((\svcs{C}_{\idxz\idxo}(\tmarg),C_{\idxo\idxt}),(\pvc{C}_{\idxz\idxt\ps \idxo}))}
\end{align*}
attains an extremum at $\tmarg=0$.

\begin{mylem}[Extremum of the KLD in \autoref{KLex}]
\label{not_global_ex}
Let $C_{1:3}$ be given as in \autoref{KLex}.
For $u_1\in(0,1)$, we define
\begin{align*}
\hdef{u_\idxz} & := \int_0^1 
\partial_{\tmarg}\svcs{F}_{\idxz|\idxo}
(u_\idxz|u_\idxo;\tmarg)\Big|_{\tmarg=0}g(u_\idxo)\text{d}u_\idxo,
\\
\Kdef{u_\idxz} & :=  
\frac{1}{\hdef{u_\idxz}}\int_0^1
\int_0^1
\partial_{\tmarg}\log \svcs{c}_{\idxz\idxt\ps \idxo}\big(\svcs{F}_{\idxz|\idxo}
(u_\idxz|u_\idxo;\tmarg),u_\idxt;\tpar\big)\Big|_{\tmarg=0}
c_{\idxlist}(u_{\idxlist})
 \text{d}u_{\idxo}  \text{d}u_{\idxt}.
\end{align*}
Then, $\forall u_\idxz\in(0,0.5)\colon \Kdef{0.5+u_\idxz}> 0\eq \tpar> 0,$ and 
$\KL{C_{\idxlist}}{\svcs{C}_{\idxz\idxo}(\tmarg),C_{\idxo\idxt},\pvc{C}_{\idxz\idxt\ps \idxo}}$
has an extremum at $\tmarg = 0$ if and only if 
\begin{align}
\label{ex1KL}
\partial_{\tmarg}\KL{C_{\idxlist}}{\svcs{C}_{\idxz\idxo}(\tmarg),
C_{\idxo\idxt},\pvc{C}_{\idxz\idxt\ps \idxo}}\Big|_{\tmarg = 0}
& = 
\int_0^{0.5}\Kdef{0.5+u_\idxz}[\hdef{0.5+u_\idxz}-\hdef{0.5-u_\idxz}]\text{d}u_\idxz = 0.
\end{align}
\end{mylem}
\begin{myproof}
See \appref{Proof_not_global_ex}.
\end{myproof}

It depends on the data generating process whether the condition in \autoref{not_global_ex} is satisfied and $\KL{C_{\idxlist}}
{\svcs{C}_{\idxz\idxo}(0),C_{\idxo\idxt},\pvc{C}_{\idxz\idxt\ps \idxo}}$ is an extremum or not as we illustrate in the following. 
If $\tpar=0$, then $\Kdef{u_\idxz}=0$ for all $u_\idxz\in(0,1)$, or if $g$ does not depend on $u_\idxo$, then $\hdef{u_\idxz}=0$ for all $u_\idxz\in(0,1)$.
Thus, the integrand in \eqref{ex1KL} is zero and we have an extremum if one of these conditions is true.
Assuming $\tpar\neq 0$ and that $g$ depends on $u_\idxo$, we see from \eqref{ex1KL} that $g$ and $\svcs{C}_{\idxz\idxo}$ determine whether we have an extremum at $\tmarg=0$. 
Depending on the copula family that is chosen for $\svcs{C}_{12}$, it may be  possible that the copula family alone determines whether $\KL{C_{\idxlist}}
{\svcs{C}_{\idxz\idxo}(0),C_{\idxo\idxt},\pvc{C}_{\idxz\idxt\ps \idxo}}$ is an extremum. 
For instance, if $\svcs{C}_{\idxz\idxo}$ is a FGM copula we obtain
\begin{align*}
\hdef{u_\idxz} & = u_1(1-u_1)\int_0^{1}(1-2u_2)g(u_2)\text{d}u_2
\shortintertext{so that}
\hdef{0.5+u_\idxz} & = \hdef{0.5-u_\idxz}, \quad \forall u_\idxz\in(0,0.5).
\end{align*}
This symmetry of $h$ across 0.5 implies that \eqref{ex1KL} is satisfied for all functions $g$. 

If we do not impose any constraints on the bivariate copulas in the first tree of the \svca{}, then
$\KL{C_{\idxlist}}{\svcs{C}_{\idxz\idxo}(0),C_{\idxo\idxt},\pvc{C}_{\idxz\idxt\ps \idxo}}$
may not even be a local minimizer of the KLD.
For instance, if $\svcs{C}_{\idxz\idxo}$ is the asymmetric FGM copula given in \eqref{asyfgm}, we find that
\begin{align*}
\hdef{u_\idxz} & = u_1^2(1-u_1)\int_0^{1}(1-2u_2)g(u_2)\text{d}u_2.
\end{align*}
If $\Lambda:=\int_0^{1}(1-2u_2)g(u_2)\text{d}u_2\neq 0$, e.g., $g$ is a non-negative function which is increasing, say $g(u_2) = u_2$, 
then, depending on the sign of $\Lambda$, either
\begin{align*}
\hdef{0.5+u_\idxz} & > \hdef{0.5-u_\idxz}, \quad \forall u_\idxz\in(0,0.5),
\shortintertext{or}
\hdef{0.5+u_\idxz} & < \hdef{0.5-u_\idxz}, \quad \forall u_\idxz\in(0,0.5),
\end{align*}
so that the integrand in \eqref{ex1KL} is either {strictly positive or negative} and thus 
$\KL{C_{\idxlist}}{C_{\idxz\idxo},C_{\idxo\idxt},\pvc{C}_{\idxz\idxt\ps \idxo}}$ 
can not be an extremum. 
Since $\theta_{12}\in[-1,1]$, it follows that 
$\KL{C_{\idxlist}}{\svcs{C}_{\idxz\idxo}(0),C_{\idxo\idxt},\pvc{C}_{\idxz\idxt\ps \idxo}}$
is not a local minimum.
As a result, we can, relating to the \PVCA{}, further decrease the KLD from the true copula if we adequately specify ``wrong'' copulas in the first tree and choose the first-order partial copula in the second tree of the \svca{}.

\subsection{Proof of \autoref{not_global_ex}}\label{Proof_not_global_ex}
The KLD attains an extremum if and only if the negative cross entropy attains an extremum. 
The negative cross entropy is given by
\begin{align*}
H_{\idxlist}({C_{\idxlist}}||{\svcs{C}_{\idxlist}}(\tmarg;\tpar)) 
& = \mathbb{E}[\log \svcs{c}_{\idxlist}(U_{\idxlist};\tmarg,\tpar)]
\\ & = 
\mathbb{E}[\log [\svcs{c}_{\idxz\idxo}(U_{\idxz:\idxo};\tmarg)
\svcs{c}_{\idxz\idxt\ps \idxo}
\big(\svcs{F}_{\idxz|\idxo}(U_\idxz|U_\idxo;\tmarg),U_\idxt;\tpar\big)]]
\\
& = \mathbb{E}[\log \svcs{c}_{\idxz\idxo}(U_{\idxz:\idxo};\tmarg)] 
+ \mathbb{E}[\log \svcs{c}_{\idxz\idxt\ps \idxo}
\big(\svcs{F}_{\idxz|\idxo}(U_\idxz|U_\idxo;\tmarg),U_\idxt;\tpar\big)].
\end{align*}
If the negative cross entropy attains an extremum then the derivative of 
$\mathbb{E}[\log \svcs{c}_{\idxlist}(U_{\idxlist};\tmarg)]$ w.r.t. $\tmarg$ is zero.
Since $\svcs{c}_{\idxz\idxo}(u_1,u_2;\tmarg)$ and $\partial_{\tmarg}\svcs{c}_{\idxz\idxo}(u_1,u_2;\tmarg)$ are both continuous on $(u_1,u_2,\tmarg)\in(0,1)^2\times\Theta_{12}$, we can apply Leibniz's rule for differentiation under the integral sign to conclude that 
$\parderiv {\mathbb{E}[\log \svcs{c}_{\idxz\idxo}(U_{\idxz:\idxo};\tmarg)]}{\tmarg}\big|_{\tmarg=0} =  {\mathbb{E}[\partial_{\tmarg}\log \svcs{c}_{\idxz\idxo}(U_{\idxz:\idxo};\tmarg)\big|_{\tmarg=0}]} = 0$ because $\svcs{C}_{\idxz\idxo}(0)$ is the true copula of $U_{\idxz:\idxo}$. 
Thus, the derivative evaluated at $\tmarg=0$ becomes
\begin{align*}
\parderiv{\mathbb{E}[\log \svcs{c}_{\idxlist}(U_{\idxlist};\tmarg,\tpar)] }{\tmarg}\Big|_{\tmarg=0}
& = \parderiv{\mathbb{E}[\log \svcs{c}_{\idxz\idxo}(U_{\idxz:\idxo};\tmarg)]}{\tmarg}\Big|_{\tmarg=0}
\\
& \quad + \parderiv{\mathbb{E}[\log \svcs{c}_{\idxz\idxt\ps \idxo}
\big(\svcs{F}_{\idxz|\idxo}(U_\idxz|U_\idxo;\tmarg),U_\idxt;\tpar\big)]}{\tmarg }\Big|_{\tmarg=0}
\\
& = \parderiv{\mathbb{E}[\log \svcs{c}_{\idxz\idxt\ps \idxo}
\big(\svcs{F}_{\idxz|\idxo}(U_\idxz|U_\idxo;\tmarg),U_\idxt;\tpar\big)]}{\tmarg }\Big|_{\tmarg=0}
\\
& = \parderiv{
\int_{[0,1]^3} \log \svcs{c}_{\idxz\idxt\ps \idxo}\big(\svcs{F}_{\idxz|\idxo}
(u_\idxz|u_\idxo;\tmarg),u_\idxt;\tpar\big) c_{\idxlist}(u_{\idxlist}) \text{d}u_{\idxlist}
}{\tmarg}\Big|_{\tmarg=0}
\\
 = 
\int_{[0,1]^3} &\frac{\parderiv{\svcs{c}_{\idxz\idxt\ps \idxo}\big(
\svcs{F}_{\idxz|\idxo}
(u_\idxz|u_\idxo;\tmarg),u_\idxt;\tpar\big)}{1}}
{\svcs{c}_{\idxz\idxt\ps \idxo}\big(\svcs{F}_{\idxz|\idxo}
(u_\idxz|u_\idxo;\tmarg),u_\idxt;\tpar\big)}
\parderiv{\svcs{F}_{\idxz|\idxo}
(u_\idxz|u_\idxo;\tmarg)}{\tmarg} \Big|_{\tmarg=0}
c_{\idxlist}(u_{\idxlist}) \text{d}u_{\idxlist},
\end{align*}
where $\parderiv{\svcs{c}_{\idxz\idxt\ps \idxo}(u,v;\tpar)}{1}$ is the partial derivative w.r.t. $u$ and
we have used Leibniz's integral rule to perform the differentiation under the integral sign for the second last equality which is valid since the integrand and its partial derivative w.r.t. $\tmarg$ are both continuous in $u_{\idxlist}$ and $\tmarg$ on $(0,1)^3\times(-1,1)$.

To compute the integral we observe that
\begin{align*}
\parderiv{\svcs{c}_{\idxz\idxt\ps \idxo}(u,v;\tpar)}{1} & = -2{\tpar}(1-2v),
\\
\parderiv{\svcs{c}_{\idxz\idxt\ps \idxo}\big(\svcs{F}_{\idxz|\idxo}
(u_\idxz|u_\idxo;0),u_\idxt;\tpar\big)}{1} & = -2\tpar(1-2u_{\idxt}),
\\
\frac{\parderiv{\svcs{c}_{\idxz\idxt\ps \idxo}\big(
\svcs{F}_{\idxz|\idxo}
(u_\idxz|u_\idxo;\tmarg),u_\idxt;\tpar\big)}{1}}
{\svcs{c}_{\idxz\idxt\ps \idxo}\big(\svcs{F}_{\idxz|\idxo}
(u_\idxz|u_\idxo;\tmarg),u_\idxt;\tpar\big)}\Big|_{\tmarg=0}
& = \frac{-2\tpar(1-2u_{\idxt})}{1+\tpar(1-2u_\idxz)(1-2u_\idxt)} =:\mdef.
\end{align*}
Note that $\mdef$ does not depend on $u_\idxo$.
Moreover,
with $c_{1:3}(u_{1:3}) = 1+g(u_2)(1-2u_1)(1-2u_3)$,
\begin{align*}
\int_0^1 \parderiv{\svcs{F}_{\idxz|\idxo}
(u_\idxz|u_\idxo;\tmarg,\tpar)}{\tmarg} \Big|_{\tmarg=0}
c_{\idxlist}(u_{\idxlist}) \text{d}u_{\idxo}
& = 
\int_0^1 \parderiv{\svcs{F}_{\idxz|\idxo}
(u_\idxz|u_\idxo;\tmarg)}{\tmarg}\Big|_{\tmarg=0}  \text{d}u_{\idxo}
\\
& \quad + (1-2u_{\idxz})(1-2u_\idxt)\int_0^1 \parderiv{
\svcs{F}_{\idxz|\idxo}
(u_\idxz|u_\idxo;\tmarg)}{\tmarg}\Big|_{\tmarg=0}g(u_\idxo)  \text{d}u_{\idxo}
\\
& = (1-2u_{\idxz})(1-2u_\idxt)\hdef{u_\idxz},
\end{align*}
where the second equality follows because
\begin{align*}
\int_0^1 \parderiv{\svcs{F}_{\idxz|\idxo}
(u_\idxz|u_\idxo;\tmarg)}{\tmarg}  \text{d}u_{\idxo}
& = \parderiv{\int_0^1 \svcs{F}_{\idxz|\idxo}
(u_\idxz|u_\idxo;\tmarg)}{\tmarg}  \text{d}u_{\idxo}
= \parderiv{u_{\idxz}}{\tmarg} = 0.
\end{align*}
Thus, integrating out $u_\idxo$, we obtain
\begin{align}
\notag
\parderiv{\mathbb{E}[\log \svcs{c}_{\idxlist}(U_{\idxlist};\tmarg)] }{\tmarg}\Big|_{\tmarg=0}
& = 
\int_{[0,1]^2} \mdef(1-2u_{\idxz})(1-2u_\idxt)\hdef{u_\idxz} \text{d}u_{\idxz}\text{d}u_{\idxt}
\\
& = \int_{[0,1]^2} \fdef{u_\idxz}{u_\idxt} \hdef{u_\idxz} \text{d}u_{\idxz}
\text{d}u_{\idxt},
\label{star}
\end{align}
where $\fdef{u_\idxz}{u_\idxt} := \mdef(1-2u_{\idxz})(1-2u_\idxt)$. We note that $\forall u_\idxz\in(0,0.5),u_\idxt\in(0,1)$:
\begin{align*}
\fdef{0.5+u_\idxz}{u_\idxt} &> 0 \eq \tpar>0,
\\
\fdef{0.5-u_\idxz}{u_\idxt} & = -\fdef{0.5+u_\idxz}{1-u_\idxt}.
\end{align*}
So, if $u_\idxz\in(0,0.5)$ then
\begin{align*}
\int_0^1 \fdef{0.5-u_\idxz}{u_\idxt}\text{d}u_\idxt & 
= {\int_0^1 \fdef{0.5-u_\idxz}{1-u_\idxt}\text{d}u_\idxt}
= -\int_0^1 \fdef{0.5+u_\idxz}{u_\idxt}\text{d}u_\idxt.
\end{align*}
Thus, if we define $\Kdef{u_\idxz}:=\int_0^1\fdef{u_\idxz}{u_\idxt}\text{d}u_\idxt$ we have that $\forall u_\idxz\in(0,0.5)$:
\begin{align}
\notag
\Kdef{0.5+u_\idxz} &  >0 \eq \tpar>0,
\\
 \Kdef{0.5-u_\idxz}&=-\Kdef{0.5+u_\idxz}.
\label{starstar}
\end{align}
Plugging this into our integral $\eqref{star}$ yields
\begin{align*}
\parderiv{\mathbb{E}[\log \svcs{c}_{\idxlist}(U_{\idxlist};\tmarg,\tpar)] }{\tmarg}\Big|_{\tmarg=0}
& =
\int_0^1\int_0^1\fdef{u_\idxz}{u_\idxt}\hdef{u_\idxz} \text{d}u_\idxz \text{d}u_\idxt
\\
& = \int_0^{0.5}\hdef{0.5-u_\idxz}\left(\int_0^1\fdef{0.5-u_\idxz}{u_\idxt}\text{d}u_{\idxt}\right)
\text{d}u_\idxz
\\
& \quad+
\int_0^{0.5}\hdef{0.5+u_\idxz}\left(\int_0^1\fdef{0.5+u_\idxz}{u_\idxt}\text{d}u_{\idxt}\right)
\text{d}u_\idxz
\\
& = \int_0^{0.5}\hdef{0.5-u_\idxz}\Kdef{0.5-u_\idxz}\text{d}u_\idxz
\\
& \quad +\int_0^{0.5}\hdef{0.5+u_\idxz}
\Kdef{0.5+u_\idxz}\text{d}u_{\idxz}
\\
& \stackrel{\eqref{starstar}}{=}
\int_0^{0.5}\Kdef{0.5+u_\idxz}[\hdef{0.5+u_\idxz}-\hdef{0.5-u_\idxz}]\text{d}u_\idxz.
\end{align*}
Note that if $\tpar=0$, then $\Kdef{u_\idxz}=0$ for all $u_\idxz\in(0,1)$, or if $g$ does not depend on $u_\idxo$, then $\hdef{u_\idxz}=0$ for all $u_\idxz\in(0,1)$, so in both cases the integrand is zero and we have an extremum.

\subsection{Proof of \autoref{non_consistency}}
\label{proof_convergence2pvc}
\autoref{non_consistency} (i) and (ii) follow directly from Theorem 1 in \citet{Spanhel2016b}, which states the asymptotic distribution of approximate rank Z-estimators if the data generating process is not nested in the parametric model family.
\autoref{non_consistency} (iii) follows then from \autoref{prop_kl} and \autoref{hop}.

\subsection{An example where the difference between $\hat{\theta}^S$ and $\hat{\theta}^J$ is more pronounced}
\label{convergence_appendix}
\begin{myex}
\label{exD}
Let $C^{\text{BB1}}(\theta,\delta)$ denote the BB1 copula with dependence parameter $(\theta,\delta)$ and $C^{\text{Sar}}(\alpha)$ be the Sarmanov copula with cdf
  $C(u,v;\alpha) = uv\big(1+(3\alpha+5\alpha^2\prod_{i=u,v}(1-2i))\prod_{i=u,v}(1-i)\big)$
for $|\alpha|\leq \sqrt{7}/5$. 
The partial Sarmanov copula  is given by 
$C^{\text{P-Sar}}(u,v;a,b) =\  uv\big(1+(3a+5b\prod_{i=u,v}(1-2i))\prod_{i=u,v}(1-i)\big)$,
where $|a|\leq \sqrt{7}/5$ and $a^2\leq b \leq (\sqrt{1-3a^2}+1)/5 $.
Define $S(u_2) = (1+\exp(u_2))^{-1}$ and $f(u_2) = 1-2S(10u_2-0.5))+ 2(1-2u_2)S(-5)$ so that $g(u_2) = 0.1\left(\sqrt{7}+1\right)\left(1-f(u_2)\right)-0.2$.
Let 
$C_{1:3}$ be the true copula with $(C_{12},C_{23},C_{13\ps 2}) 
= (C^{\text{BB1}}(2,2),C^{\text{BB1}}(2,2),C^{{\text{Sar}}}\!(g(u_2))$  and 
$\svcs{C}_{1:3} = 
(C^{\text{BB1}}(2,2),C^{\text{BB1}}(2,2),C^{{\text{P-Sar}}}(a,b))$ be the parametric \SVCM{} that is fitted to data generated from $C_{1:3}$.
\end{myex}

\begin{figure}[h!]
\centering
\begin{subfigure}[t]{0.5\textwidth}
  \centering
    \includegraphics[scale=\scaleit]{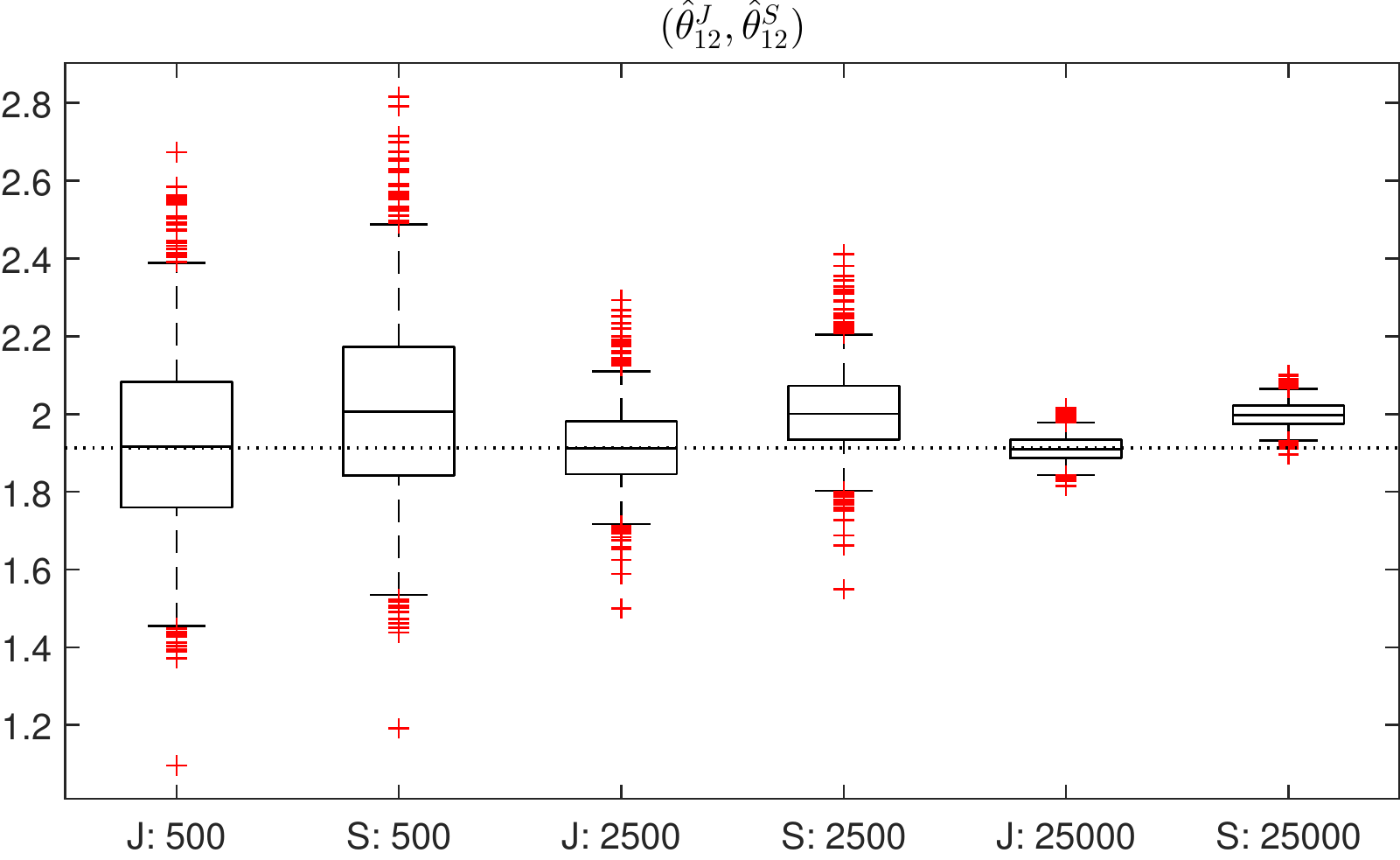}
\end{subfigure}%
\begin{subfigure}[t]{0.5\textwidth}
  \centering
    \includegraphics[scale=\scaleit]{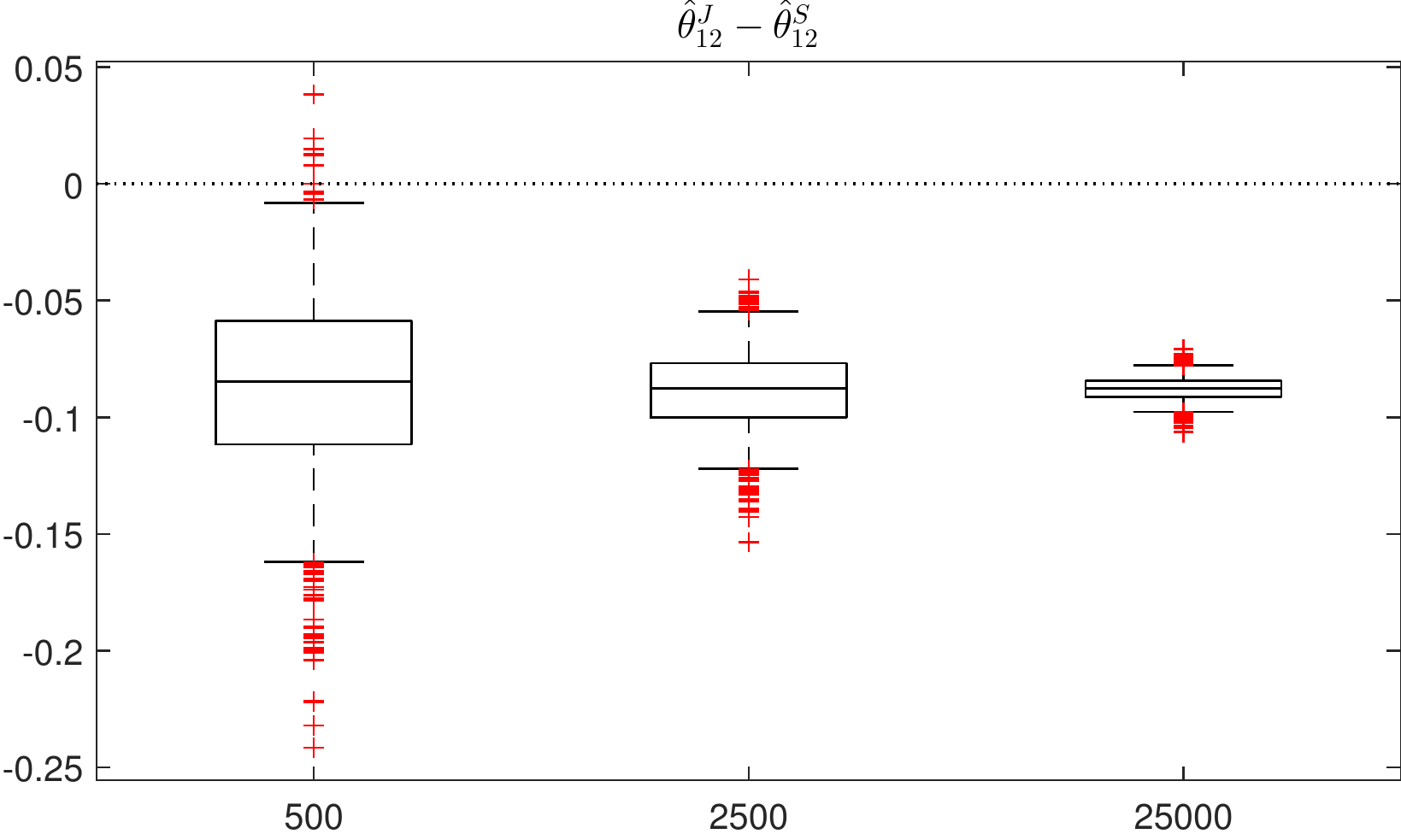}
\end{subfigure}
\medbreak

\begin{subfigure}[t]{0.5\textwidth}
  \centering
    \includegraphics[scale=\scaleit]{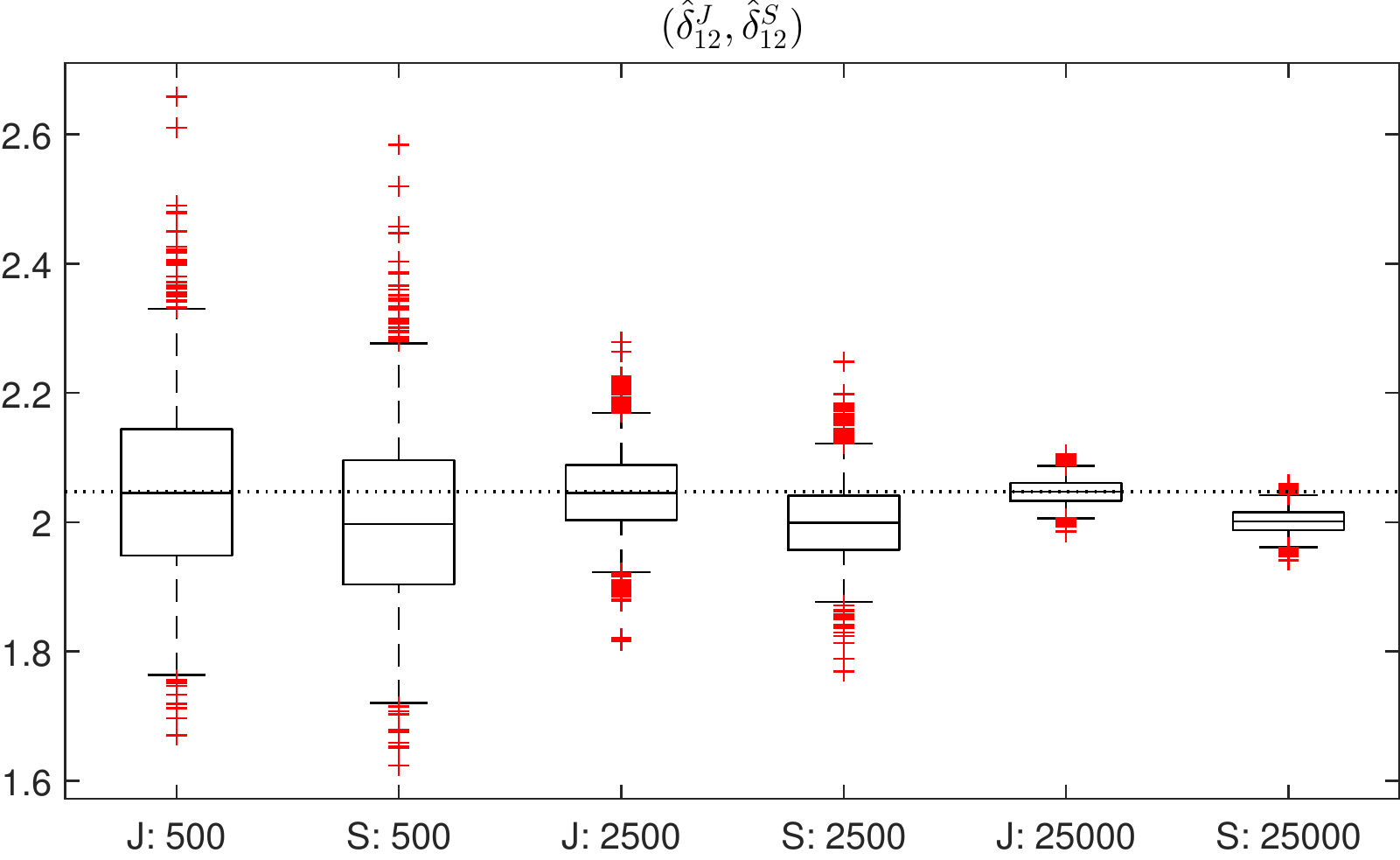}
\end{subfigure}%
\begin{subfigure}[t]{0.5\textwidth}
  \centering
    \includegraphics[scale=\scaleit]{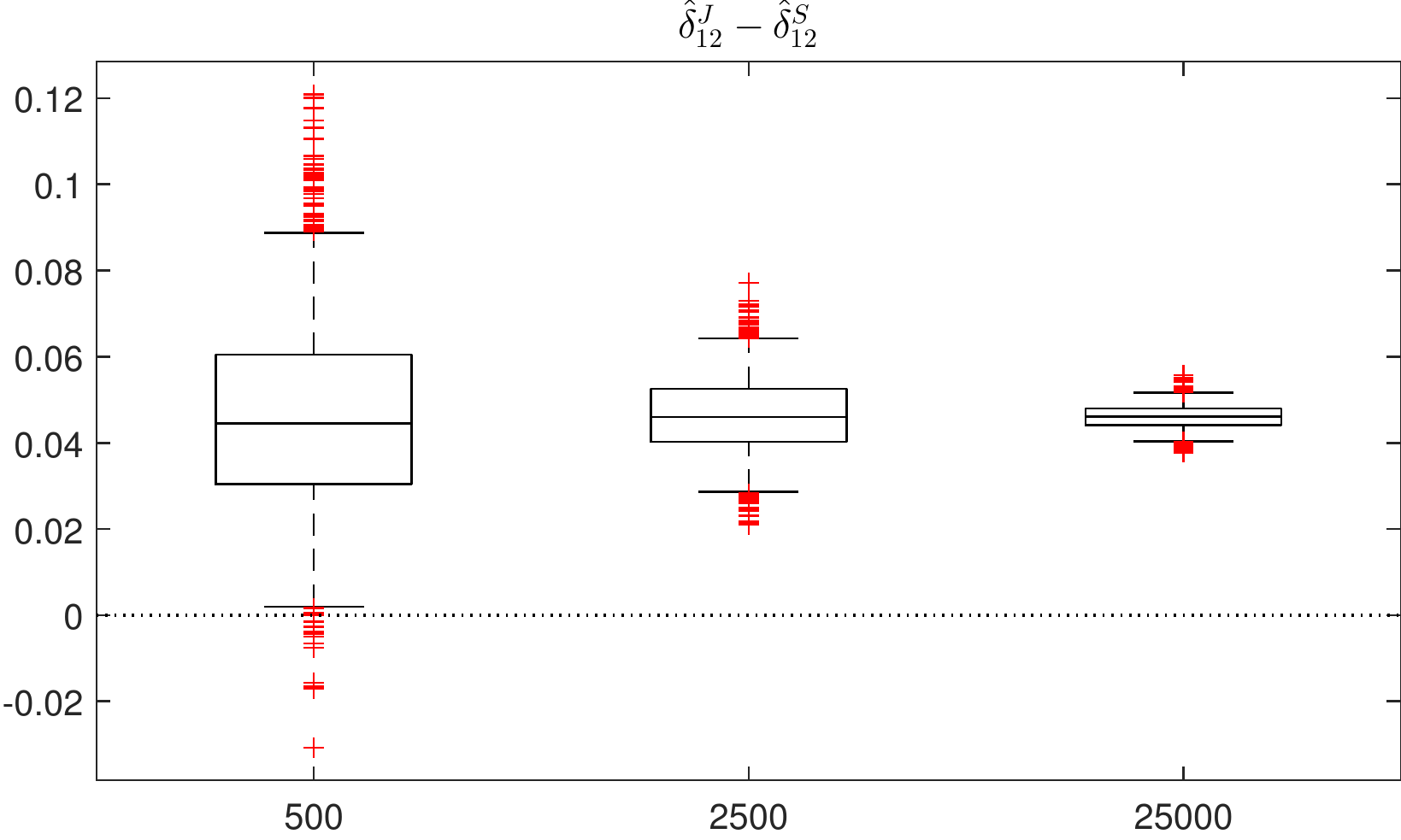}
\end{subfigure}
\medbreak

\begin{subfigure}[t]{0.5\textwidth}
  \centering
    \includegraphics[scale=\scaleit]{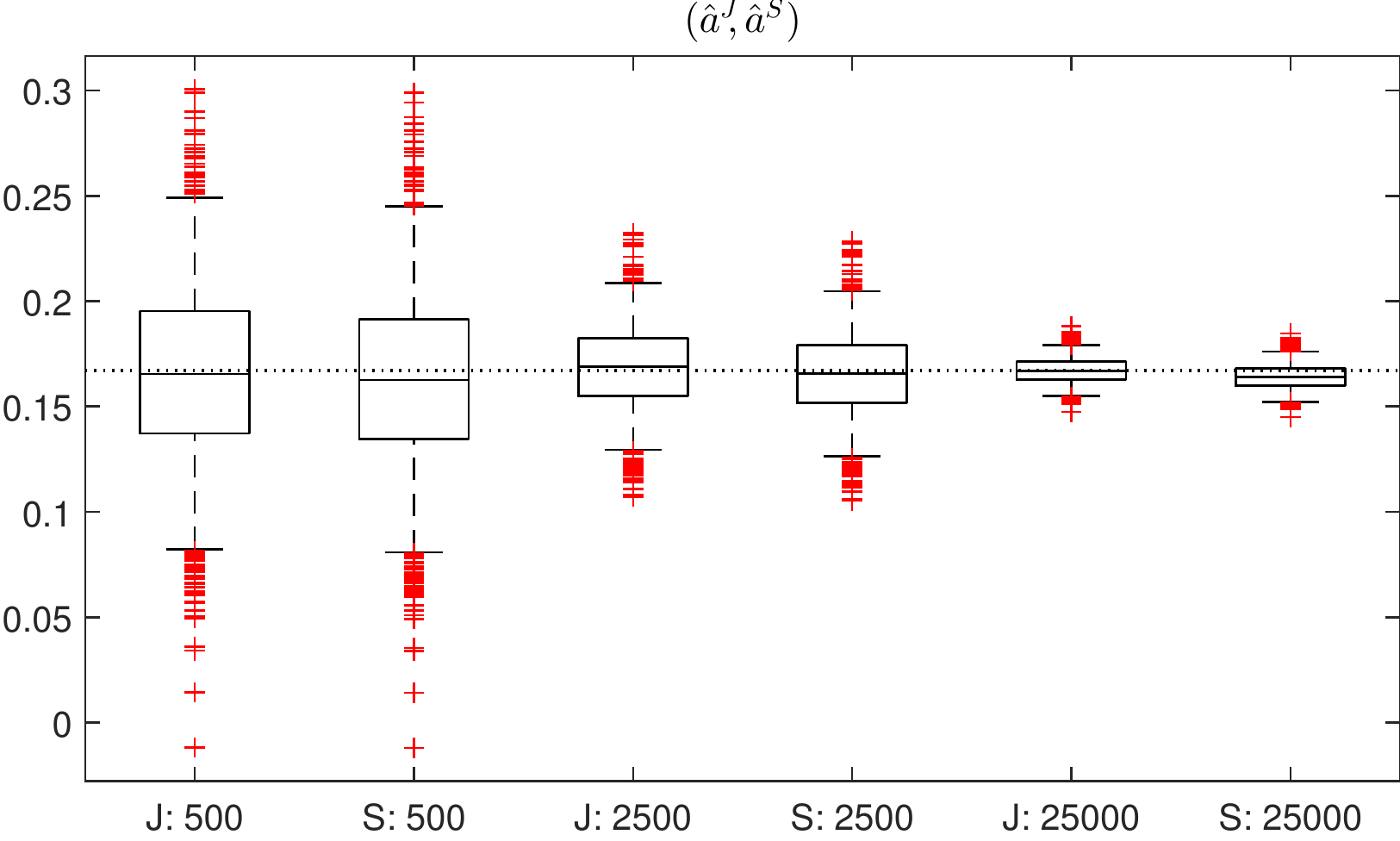}
\end{subfigure}%
\begin{subfigure}[t]{0.5\textwidth}
  \centering
    \includegraphics[scale=\scaleit]{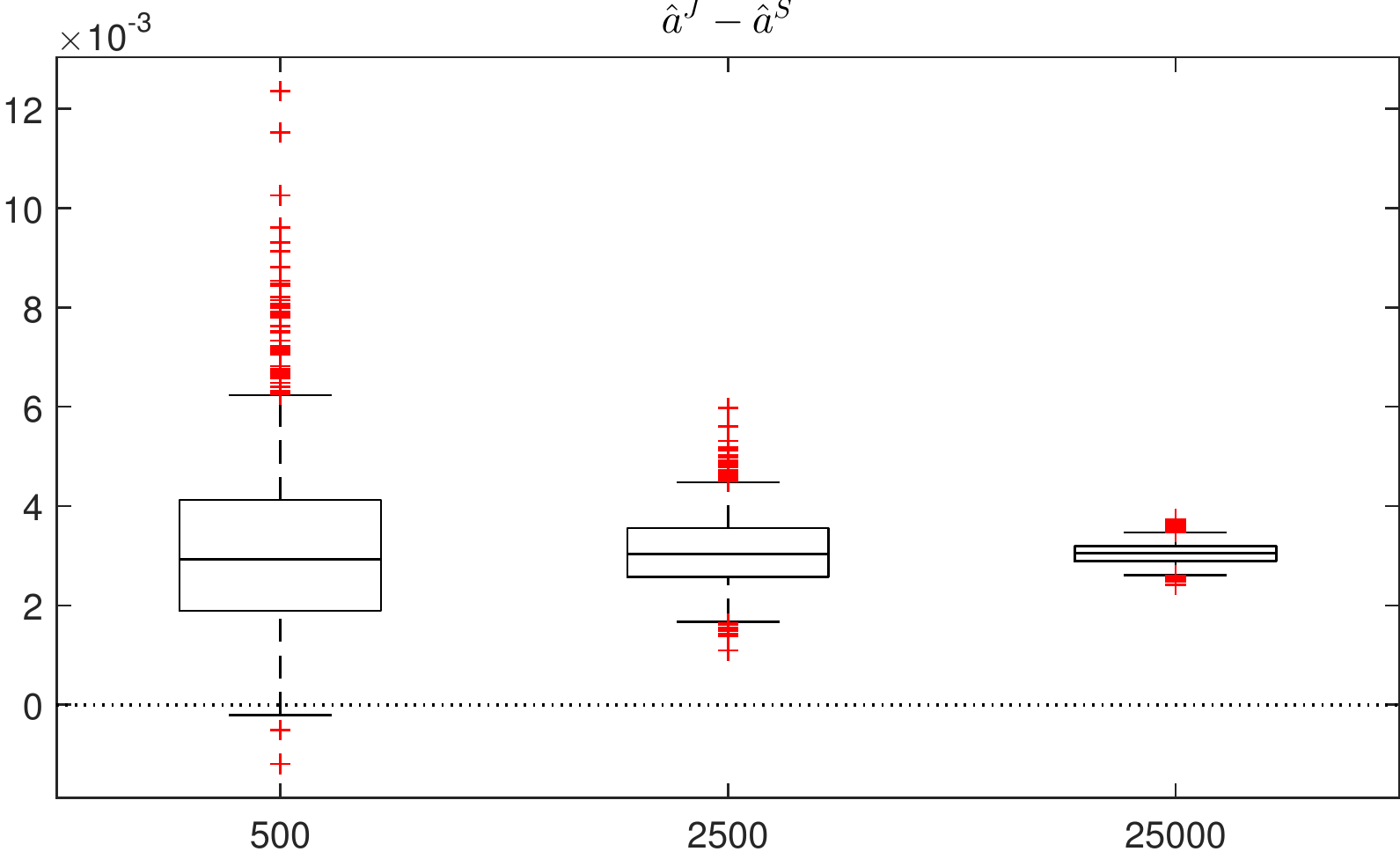}
\end{subfigure}
\medbreak

\begin{subfigure}[t]{0.5\textwidth}
  \centering
    \includegraphics[scale=\scaleit]{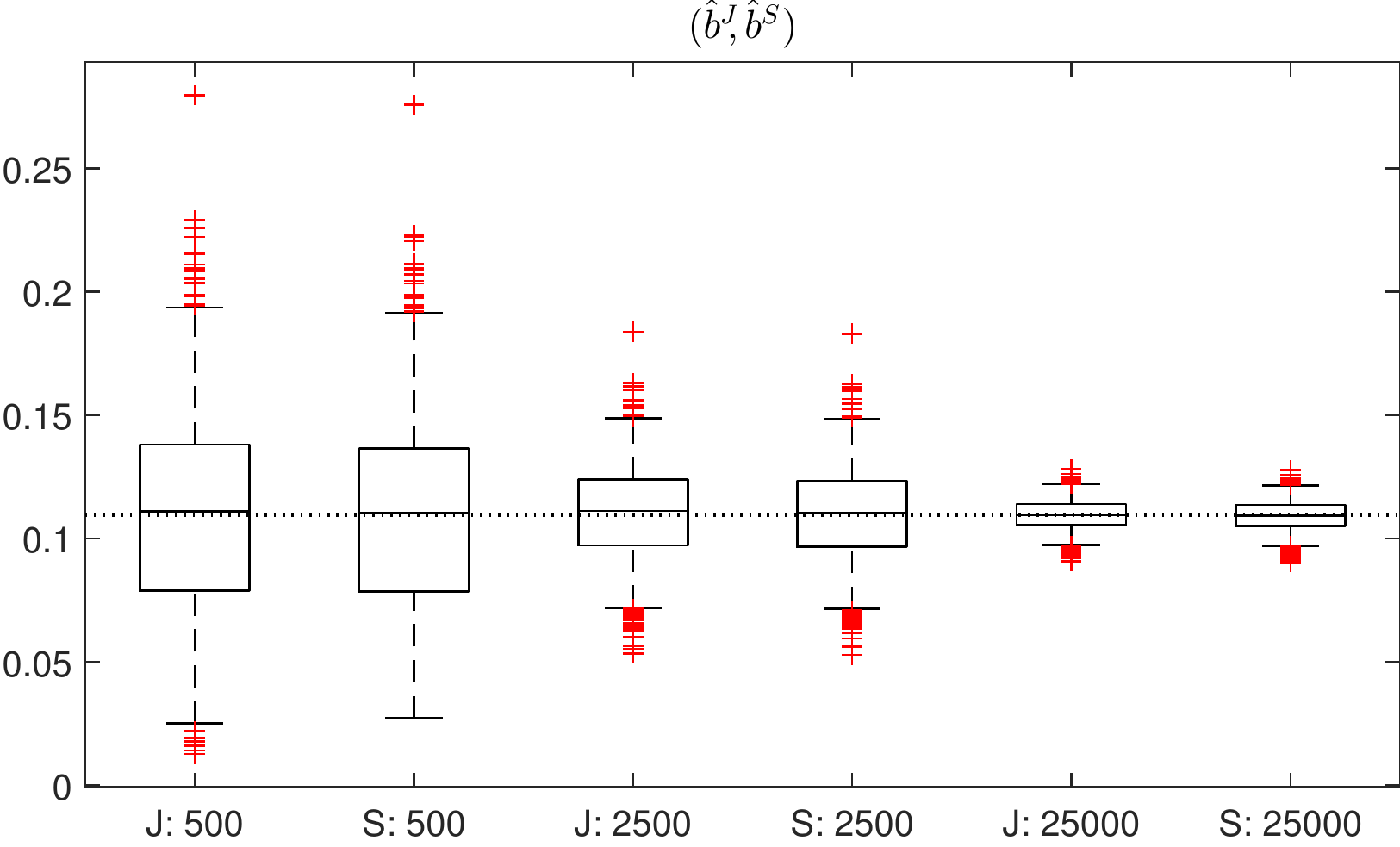}
\end{subfigure}%
\begin{subfigure}[t]{0.5\textwidth}
  \centering
    \includegraphics[scale=\scaleit]{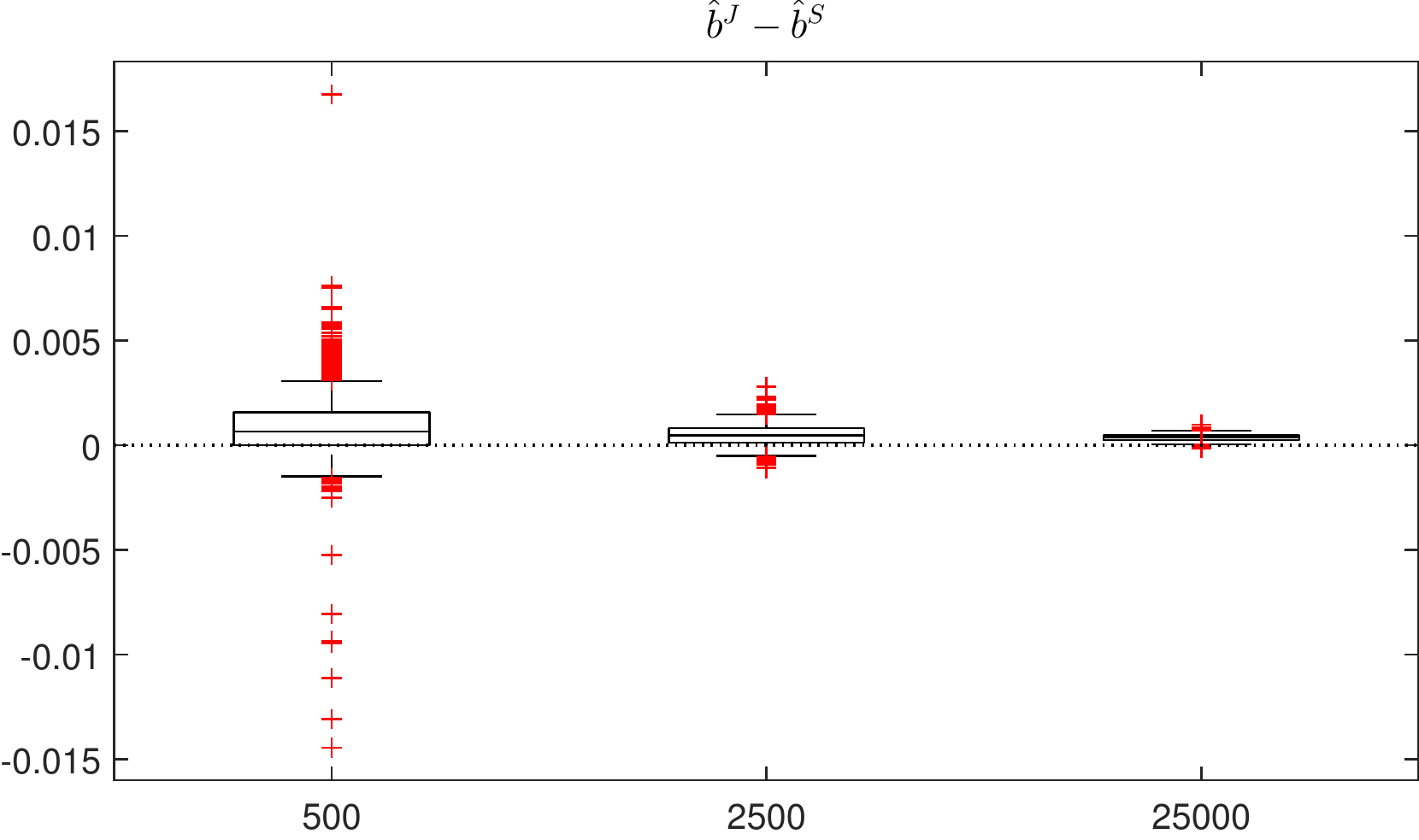}
\end{subfigure}

\captitleno{exD}{pseudo-true parameter}
\label{figexD}

\end{figure}

Note that $g$ is a sigmoid function,
with $(g(0),g(1)) = (-0.2,\sqrt{7}/5)$, so that Spearman's rho of the conditional copula $C^{{\text{Sar}}}\!(g(u_2))$ varies in the interval $(g(0),g(1)) = (-0.2,\sqrt{7}/5)$ because $\rho_{C^{\text{Sar}}} = \alpha$.
\autoref{figexD} shows that the difference between step-by-step and joint ML estimates for the two parameters of 
the first copula in the first tree is already (individually) significant at the 5\% level if the sample size is 500 observations.
Thus, the difference between step-by-step and joint ML estimates can be relevant for moderate sample sizes if the variation
in the conditional copula is strong enough.  
Once again, the difference between step-by-step and joint ML estimates is less pronounced for the parameters of $\svcs{C}_{13;2}$ but it
also becomes highly significant with sufficient sample size.

\end{document}